\documentclass[twocolumn,showpacs,showkeys,prc,aps,amsmath,amssymb,nofootinbib,floatfix,10pt]{revtex4-1}

\usepackage{graphicx}
\usepackage{color}
\usepackage{multirow}
\usepackage{bm}% bold math
\usepackage{mathtools}
\usepackage[caption=false]{subfig}
\usepackage{url}
\usepackage{breakurl}
\usepackage{physics}
\usepackage{cancel}
\usepackage{hyperref}
\usepackage{bibentry}
\usepackage{slashed}
\usepackage{simplewick}
\allowdisplaybreaks

\DeclareMathAlphabet{\mathcal}{OMS}{cmsy}{m}{n}
\DeclareMathSizes{12}{12}{8}{5}
\newcommand{\rf}[1]{Ref.~\cite{#1}}
\newcommand{\eq}[1]{Eq.~(\ref{#1})}
\newcommand{\fg}[1]{Fig.~\ref{#1}}
\newcommand{\non}{\nonumber \\}
\newcommand{\vc}[1]{\mathbf{#1}}
\renewcommand{\varepsilon}{\text{\usefont{OML}{cmr}{m}{n}\symbol{15}}}
\def\<{\langle}
\def\>{\rangle}
\def\({\Big{(}}
\def\){\Big{)}}
\def\[{\Big{[}}
\def\]{\Big{]}}
\def\nn{\nonumber}
\def\gm{\gamma^{\mu}}

\def\gf{\gamma^5}

\def\NDBD{0\nu\beta\beta}
\def\TNDBD{2\nu\beta\beta}
\def\Q{Q_{\beta\beta}}
\def\effmass{\<m_{\beta\beta}\>}
\begin{document}

\title{Interference Effects for $\NDBD$ Decay in the Left-Right Symmetric Model}
\author{Fahim Ahmed}
\email{ahmed1f@cmich.edu}
\author{Mihai Horoi}
\email{mihai.horoi@cmich.edu}
\affiliation{Department of Physics, Central Michigan University, Mount Pleasant, Michigan 48859, USA}%
\date{\today}

%%%%%%%%%%%%%%%%%%%%%%%%%%%%%%%%%%%%%%%%%%%%%%%%%%%%%%%%%%%%%%%%%%%%%%%%%%%%%%%%%%%%%%%%%%%%%%%%%%%%
\begin{abstract}
Various mechanisms may contribute to neutrinoless double beta decay in the left-right symmetric model. The interference between these mechanisms also contribute to the overall decay rate. The analysis of the contributions of these interference terms is important for disentangling different mechanisms. In the present paper we study interference effects contributing to the decay rate for neutrinoless double-$\beta$ decay in the left-right symmetric model. The numerical values for maximum interference for several nuclides are calculated. It is observed that, for most of the interference terms, the contribution is smaller than 20$\%$ for all the nuclei considered in the study. However, the interference between the mass-mechanisms (light and heavy) and $\eta$ mechanism is observed to be in the range 30$\%$-50$\%$. The variation of the interference effect with the $Q$ values is also studied.       
\end{abstract}
%%%%%%%%%%%%%%%%%%%%%%%%%%%%%%%%%%%%%%%%%%%%%%%%%%%%%%%%%%%%%%%%%%%%%%%%%%%%%%%%%%%%%%%%%%%%%%%%%%%%

\maketitle

\renewcommand*\arraystretch{1.3}

%%%%%%%%%%%%%%%%%%%%%%%%%%%%%%%%%%%%%%%%%%%%%%%%%%%%%%%%%%%%%%%%%%%%%%%%%%%%%%%%%%%%%%%%%%%%%%%%%%%%
\section{Introduction}\label{Intro}
The lepton number violating (LNV) rare nuclear process of neutrinoless double-$\beta$ decay ($\NDBD$),
\begin{align}\label{decays}
\prescript{A}{Z}{\mathrm{X}}\rightarrow  \prescript{A}{Z+2}{\mathrm{X}}+2e^-,
\end{align}
could be an important low-energy manifestation of physics beyond the Standard Model (BSM). In contrast with the two neutrino double- $\beta$ decay ($\TNDBD$), where two antineutrinos are also emitted, in $\NDBD$ the lepton number is violated by two units ($\Delta L=2$). Experimental observation of $\NDBD$ would indicate BSM physics since lepton number is conserved in the Stardard Model (SM). In addition, $\NDBD$ would prove the Majorana nature of neutrinos \cite{PhysRevD.25.2951}. Apart from the extensively studied "standard mass-mechanism" of light left-handed (LH) neutrino exchange \cite{PhysRevC.87.014320}, several BSM mechanisms are proposed to contribute to the $\NDBD$ decay \cite{PhysRevC.87.014320, 0034-4885-75-10-106301}. 

The left-right symmetric model (LRSM) is a natural extension of the SM where the parity is assumed to be restored at energies higher than the electroweak scale. Actively investigated at the LHC \cite{Khachatryan:2014dka}, in the LRSM scenario several competing mechanisms contribute to $\NDBD$ due to the presence of the right-handed (RH) fields \cite{Barry:2013xxa}. Additionally, LRSM provides a natural framework for type-I \cite{PhysRevLett.44.912} and type-II \cite{Mohapatra+Senjanovic1981} seesaw mechanisms generating small neutrino masses. Moreover, the seesaw mechanism requires the existence of heavy, sterile neutrinos \cite{Mohapatra+Senjanovic1981}. Neutrino mixing schemes would then naturally incorporate heavy-mass eigenstates for both LH and RH neutrinos (see Sec. \ref{LRSM} for details). 

The study of the $\NDBD$ decay rate allows us to extract the new neutrino physics parameters resulting from such BSM physics scenarios. However, the neutrino oscillation experiments alone can not determine the absolute masses of the neutrinos. Moreover, if the regular "mass-mechanism" dominates, then $\NDBD$ decay will allow us to determine the absolute masses of neutrinos. All these features make $\NDBD$ an exciting process for probing BSM physics. It thus becomes essential to disentangle the competing underlying mechanisms inducing $\NDBD$  in order to extract these new neutrino physics parameters arising from BSM physics \cite{PhysRevD.83.113003}. The inverse half-life formula for $\NDBD$ has the following general structure:
\begin{align}\label{genhalf-life}
\[T_{1/2}^{0\nu}\]^{-1}=\abs{\sum_{i}\Big(\text{PPP}\Big)_i\times\Big(\text{PSF}\Big)_i^{\frac{1}{2}}\times\Big(\text{NME}\Big)_i}^2.
\end{align}
Here, PPP are the particle physics parameters arising from BSM physics, the phase-space factors (PSF) take into account the kinematical factors of the two outgoing electrons, and NME are the nuclear matrix elements for the nuclear transition between the initial and final nuclei. The summation $i$ is over all possible amplitudes that could induce the $\NDBD$ process. Because of the modulus squared, interference between different terms in \eq{genhalf-life} also contribute to the total decay rate of the process. In \rf{Ahmed:2017pqa} we studied the interference between the standard mass mechanism and heavy RH neutrino exchange mechanism. Our analysis in \cite{Ahmed:2017pqa} showed dependence of the relative interference factor on the Q value of $\NDBD$ ($\Q$). A contribution no larger than 12$\%$ was found for all the nuclei considered. Here we extend our study of interference to other relevant pairs of mechanisms, inducing $\NDBD$ in the LRSM for six nuclei of current experimental interest.

The paper is organized as follows: Section \ref{LRSM} gives a brief outline of the LRSM followed by the general formalism for $\NDBD$ in LRSM in Sec. \ref{NDBD}. In Sec. \ref{Analysis} we present the analysis of the interference terms with the numerical results.
\vspace{-10pt}    
\section{Brief Review of the Left-Right Symmetric Model}\label{LRSM}
In the LRSM the SM gauge group $\mathcal{G}_{SM}\equiv SU(3)_C\otimes SU(2)_L\otimes U(1)_Y$ is extended to $SU(3)_C\otimes\mathcal{G}_{LR}$ with $\mathcal{G}_{LR}\equiv SU(2)_L\otimes SU(2)_R\otimes U(1)_{B-L}$ \cite{Mohapatra+Senjanovic1981, Pati:1974yy, Mohapatra:1974gc}. Restoring parity above the electroweak scale, the extended group $SU(2)_R$ allows us to form the RH fermions as doublets. We have the following fermion particle content in LRSM with the corresponding representation under $SU(3)_C\otimes\mathcal{G}_{LR}$ \cite{Barry:2013xxa, Dev:2014xea, Borah:2017ldt},     
\begin{align}
& SU(2)_L \,\,\text{Leptons}:\, L_{L j}=\left(\begin{array}{c}{\nu_{Lj}} \\ {e_{Lj}}\end{array}\right) \in (\mathbf{1},\mathbf{2},\mathbf{1},-1),\\ 
& SU(2)_R \,\,\text{Leptons}:\, L_{R j}=\left(\begin{array}{c}{\nu_{Rj}} \\ {e_{Rj}}\end{array}\right) \in (\mathbf{1},\mathbf{1},\mathbf{2},-1), \\ 
& SU(2)_L \,\,\text{Quarks}:\, Q_{L j}=\left(\begin{array}{c}{u_{Lj}} \\ {d_{Lj}^{\prime}}\end{array}\right) \in \left(\mathbf{3},\mathbf{2},\mathbf{1},\tfrac{1}{3}\right),\\ 
& SU(2)_R \,\,\text{Quarks}:\, Q_{R j}=\left(\begin{array}{c}{u_{Rj}} \\ {d_{Rj}^{\prime}}\end{array}\right) \in \left(\mathbf{3},\mathbf{1},\mathbf{2},\tfrac{1}{3}\right),
\end{align}
where the generations are defined as: $\nu_{j=1,2,3}\equiv\{\nu_e, \nu_\mu, \nu_\tau\}$, $e_{j=1,2,3}\equiv\{e, \mu, \tau \}$, $u_{j=1,2,3}\equiv\{u, c, t \}$, $d^{\prime}_{j=1,2,3}\equiv\{d^{\prime}, s^{\prime}, b^{\prime}\}$. The subscripts $L$ and $R$ are associated with the chiral projection operators $P_{L}=\tfrac{1}{2}(1-\gf)$ and $P_{R}=\tfrac{1}{2}(1+\gf)$, respectively. The first three entries of the quadruplet of numbers denote the dimension of the representation under each of the gauge groups $SU(3)_C$, $SU(2)_L$, $SU(2)_R$, respectively \cite{Burgess:2007zi}. The fourth entry denotes the quantum number associated with the group $U(1)_{B-L}$: the difference between the baryon and lepton number, $B-L=2(Q-T_{3L}-T_{3R})$, with $Q$ being the electromagnetic charge and $T_{3L}$ and $T_{3R}$ being the third component of the isospin corresponding to $SU(2)_{L}$ and $SU(2)_{L}$, respectively. For example, $(\mathbf{3},\mathbf{1},\mathbf{2},\frac{1}{3})$ for $SU(2)_R$ quarks denote a triplet under $SU(3)_C$, a singlet under $SU(2)_L$, a doublet under $SU(2)_R$, and has a charge $\frac{1}{3}$ under $U(1)_{B-L}$ \cite{Borah:2017ldt}. The seven massless gauge bosons along with their respective couplings for the $\mathcal{G}_{LR}^{EW}$ sector are, 
\begin{align}
& SU(2)_{L} : g_{L},\,\, \{W_{L \mu}^{1}, W_{L \mu}^{2}, W_{L \mu}^{3}\}, \\ 
& SU(2)_{R} : g_{R}, \,\, \{W_{R \mu}^{1}, W_{R \mu}^{2}, W_{R \mu}^{3}\},\\ 
& U(1)_{B-L} : g^{\prime}, \,\, B_{\mu}.
\end{align} 
The interaction Lagrangian before spontaneous symmetry breaking (SSB) between fermions and gauge bosons for the $\mathcal{G}_{LR}$ sector is constructed in parallel to the SM electroweak Lagrangian (a summation of repeated indices, $j,a=1,2,3$ and $\mu=0,1,2,3$, is implied),  
\begin{align} 
\mathcal{L}^{\text{\tiny{EW}}}_{\text{\tiny{LR}}}&=g_{L}\left[\overline{L}_{Lj}\gamma^{\mu}\frac{\sigma_a}{2}L_{Lj}+\overline{Q}_{Lj}\gamma^{\mu}\frac{\sigma_a}{2} Q_{Lj}\right]W_{L \mu}^{a}\non &+g_{R}\left[\overline{L}_{Rj}\gamma^{\mu}\frac{\sigma_a}{2}L_{Rj}+\overline{Q}_{Rj}\gamma^{\mu}\frac{\sigma_a}{2} Q_{Rj}\right]W^a_{R\mu}\non &+g^{\prime}\left[\overline{L}_{Lj}\gamma^{\mu}\tfrac{B-L}{2}L_{Lj}+\overline{Q}_{Lj}\gamma^{\mu}\tfrac{B-L}{2}Q_{Lj}\right.\non &\left. +\overline{L}_{Rj}\gamma^{\mu}\tfrac{B-L}{2}L_{Rj}+\overline{Q}_{Rj}\gamma^{\mu}\tfrac{B-L}{2}Q_{Rj}\right]B_{\mu}.\label{Lferm} 
\end{align}
The charge-current part of $\mathcal{L}^{\text{\tiny{EW}}}_{\text{\tiny{LR}}}$, which is relevant for $\NDBD$, takes the following form (confining ourselves to only the first generation),
\begin{align}
\mathcal{L}_{\text{\tiny{LR}}}^{\text{\tiny{CC}}}\supseteq \frac{g_{L}}{\sqrt{2}}&\[\(\overline{\nu_{eL}}\gamma^{\mu}e_{L}+\overline{u_L} \gamma^{\mu} d^{\prime}_{L}\)W_{\mu L}^{+}\non &+\(\overline{e_L}\gamma^{\mu}\nu_{eL}+\overline{d^{\prime}_{L}} \gamma^{\mu} u_{L}\)W_{\mu L}^{-}\]\non
&+\frac{g_{R}}{\sqrt{2}}\[\(\overline{\nu_{eR}}\gamma^{\mu}e_{R}+\overline{u_R} \gamma^{\mu} d^{\prime}_{R}\)W_{\mu R}^{+}\non &+\(\overline{e_R}\gamma^{\mu}\nu_{eR}+\overline{d^{\prime}_{R}} \gamma^{\mu} u_{R}\)W_{\mu R}^{-}\],
\end{align}
where the charged vector bosons are defined in terms of the $W_{L(R)\mu}^{a=1,2,3}$ fields as,
\begin{align}
& W_{L(R)\mu}^{\pm}=\frac{1}{\sqrt{2}}\(W_{L(R)\mu}^{1}\mp i W_{L(R)\mu}^{2}\).
\end{align}  
The scalar sector consists of two Higgs triplets and a bi-doublet \cite{Grimus:1993fx}, 
\begin{align}
\Delta_{L(R)}=\left[\begin{array}{cc}{\frac{1}{\sqrt{2}}\Delta^{+}_{L(R)}} & {\Delta_{L(R)}^{++}} \\ {\Delta_{L(R)}^{0}} & {\frac{-1}{\sqrt{2}}\Delta_{L(R)}^{+}}\end{array}\right]\, ,\,
\Phi =\left[\begin{array}{cc}{\phi_{1}^{0}} & {\phi_{2}^{+}} \\ {\phi_{1}^{-}} & {\phi_{2}^{0}}\end{array}\right],
\end{align}
with $\Delta_{L} \in (1,\vc{3},1,2)$, $\Delta_{R} \in (1, 1,\vc{3},2)$ and $\Phi \in (1,\vc{2},\vc{2},0)$. The gauge symmetry $\mathcal{G}_{LR}$ is broken in two stages by the scalar sector of the theory. Above the SM electroweak scale the SSB: $SU(2)_{L} \otimes SU(2)_{R} \otimes U(1)_{B-L} \rightarrow SU(2)_{L} \otimes U(1)_{Y}$ takes place through the vaccum expectation value (VEV) of the two Higgs triplets,
\begin{align}
\left\langle\Delta_{L}\right\rangle=\left(\begin{array}{cc}{0} & {0} \\ {\frac{1}{\sqrt{2}}v_{L} e^{i \theta_{L}}} & {0}\end{array}\right)\, ,\, \left\langle\Delta_{R}\right\rangle=\left(\begin{array}{cc}{0} & {0} \\ {\frac{1}{\sqrt{2}}v_{R}} & {0}\end{array}\right),
\end{align}
This breaks the parity and also allows Majorana mass terms for neutrinos. In the second stage, the SM electroweak SSB: $SU(2)_{L} \otimes U(1)_{Y} \rightarrow U(1)_{E M}$ takes place through the VEV of the bi-doublet Higgs, 
\begin{equation}
\langle\Phi\rangle=\left(\begin{array}{cc}{\frac{1}{\sqrt{2}}\kappa_{1}} & {0} \\ {0} & {\frac{1}{\sqrt{2}}\kappa_{2} e^{i \alpha}}\end{array}\right).
\end{equation} 
Here we have written the Lagrangian in the flavor basis. After SSB $\mathcal{L}_{\text{\tiny{LR}}}^{\text{\tiny{EW}}}$ acquires mass terms for the fermions and gauge bosons. For the neutrino sector, type I + II seesaw scenario is assumed, giving rise to small masses for light neutrinos due to the presence of heavy Majorana neutrinos \cite{Barry:2013xxa}. The mass-matrix for neutrinos ($\nu_{e,\mu,\tau}$), d type quarks ($d^{\prime}_j$) and the charged vector bosons ($W_{L(R)}^{\pm}$) are not diagonal in the flavor basis. We thus reexpress the flavor-basis fields in terms of fields in the mass-basis diagonalizing the mass matrices, for $d^{\prime}$ quarks:
\begin{align}
& d_L^{\prime}=V_{ud}d_L+V_{us}s_L+V_{ub}b_L, \label{dL}\\
& d_R^{\prime}=V^{\prime}_{ud}d_R+V^{\prime}_{us}s_R+V^{\prime}_{ub}b_R, \label{dR}
\end{align}
for electron-neutrinos:
\begin{align}
& \nu_{eL}=\sum_{i=1,2,3}^{\text{light}}U_{ei}\nu_{Li}+\sum_{i=1,2,3}^{\text{heavy}}S_{ei}(N_{Ri})^c,\\
& \nu_{eR}=\sum_{i=1,2,3}^{\text{light}}T^*_{ei}(\nu_{Li})^c+\sum_{i=1,2,3}^{\text{heavy}}V^*_{ei}N_{Ri},\label{neu-mix}
\end{align}
and for $W$ bosons:
\begin{align}
\left(\begin{array}{c}{W_{L}^{ \pm}} \\ {W_{R}^{ \pm}}\end{array}\right)=\left(\begin{array}{cc}{\cos \xi} & {\sin \xi e^{i \alpha}} \\ {-\sin \xi e^{-i \alpha}} & {\cos \xi}\end{array}\right)\left(\begin{array}{c}{W_{1}^{ \pm}} \\ {W_{2}^{ \pm}}\end{array}\right).
\end{align}
Here \eq{dL} is the first row of the Cabibbo-Kobayashi-Maskawa (CKM) matrix for LH quark mixing with \eq{dR} being the first row of an equivalent CKM matrix for RH quark mixing \cite{Senjanovic:2015yea}. The matrix elements $V_{ud}$ and $V^{\prime}_{ud}$ can be approximated as $V_{ud}\simeq\cos{\theta_c}$ and $V^{\prime}_{ud}\simeq\cos{\theta_c^{\prime}}$ in terms of the Cabibbo angle $\theta_c$ for LH $d$ quarks and analogous $\theta_c^{\prime}$ for RH $d$ quarks \cite{Doi+Kotani1985}. We have considered the $(3\text{-light} + 3\text{-heavy})$ scenario for Majorana neutrino mixing wherein the mass basis of the light-neutrinos are $\nu_i$ with masses $m_i$, and the heavy neutrinos are $N_i$ with masses $M_i$. The $S, T, V$ mixing matrices are generalization of the PMNS matrix $U$ for the LH-light neutrino mixing. The charged $W$ bosons, $W^{\pm}_{L(R)}$ are linear combination of physical bosons $W^{\pm}_{1(2)}$ with definite masses $m_{W_1}$ and $m_{W_2}$, respectively. We can further assume a discrete LR symmetry where the Lagrangian is invariant under the exchange $L\leftrightarrow R$. This assumption requires that the two gauge couplings be equal, $g=g_L=g_R$. The case of $g_L\neq g_R$ leads to different expressions for the effective couplings $G_F$, $\lambda$, $\eta$ (see below), but the form of the $\NDBD$ amplitudes are the same as for $g_L=g_R$ (see \rf{Hirsch:1995rf} for details). Thus, under these assumptions we can write the charged-current Lagrangian for the first fermion generation in the mass basis as,
\begin{widetext}
\vspace{-5mm}
\begin{align}\label{Weak-L}
\mathcal{L}^{\text{\tiny{CC}}}_{\text{\tiny{LR}}} \supseteq \frac{g}{\sqrt{2}} \sum_{i=1}^{3}& \Bigg[\bigg[\(U^{*}_{ei}\overline{\nu_{Li}}+S^{*}_{ei}\overline{(N_{Ri})^c}\)\gm e_L + \cos{\theta_c}\overline{u_L}\gm d_L\bigg]\(\cos{\xi}W_{1\mu}^{+}+\sin{\xi} e^{i \alpha} W_{2\mu}^{+}\)\nn\\ 
&+\bigg[\overline{e_L}\gamma^{\mu}\(U_{ei}\nu_{Li}+S_{ei}(N_{Ri})^{c}\)+\cos{\theta_c}\overline{d_L}\gm u_L\bigg]\(\cos{\xi}W_{1 \mu}^{-}+\sin{\xi} e^{i \alpha} W_{2 \mu}^{-}\)\nn\\
&+\bigg[\(T_{ei}\overline{(\nu_{Li})^c}+V_{ei}\overline{N_{Ri}}\)\gm e_R + \cos{\theta^{\prime}_c}\overline{u_R}\gm d_R\bigg]\(-\sin{\xi} e^{-i\alpha} W_{1\mu}^{+}+ \cos{\xi} W_{2\mu}^{+}\)\nn\\ 
&+\bigg[\overline{e_R}\gamma^{\mu}\(T^*_{ei}(\nu_{Li})^c+V^*_{ei}N_{Ri}\)+\cos{\theta^{\prime}_c}\overline{d_R}\gm u_R\bigg]\(-\sin{\xi} e^{-i\alpha} W_{1 \mu}^{-}+ \cos{\xi} W_{2 \mu}^{-}\)\Bigg]. 
\end{align}
\vspace{-5mm}
\end{widetext}
%%%%%%%%%%%%%%%%%%%%%%%%%%%%%%%%%%%%%%%%%%%%%%%%%%%%%%%%%%%%%%%%%%%%%%%%%%%%%%%%%%%%%%%%%%%%%%%%%%%%

%%%%%%%%%%%%%%%%%%%%%%%%%%%%%%%%%%%%%%%%%%%%%%%%%%%%%%%%%%%%%%%%%%%%%%%%%%%%%%%%%%%%%%%%%%%%%%%%%%%%

\section{Formalism for $\NDBD$ in the Left-Right Symmetric Model}\label{NDBD}
\subsection{$\beta$-decay in left-right symmetric model}
\begin{figure*}[htp]
\centering
\subfloat[$W_L$ mediation for purely LH fields.]{\includegraphics[scale=0.27]{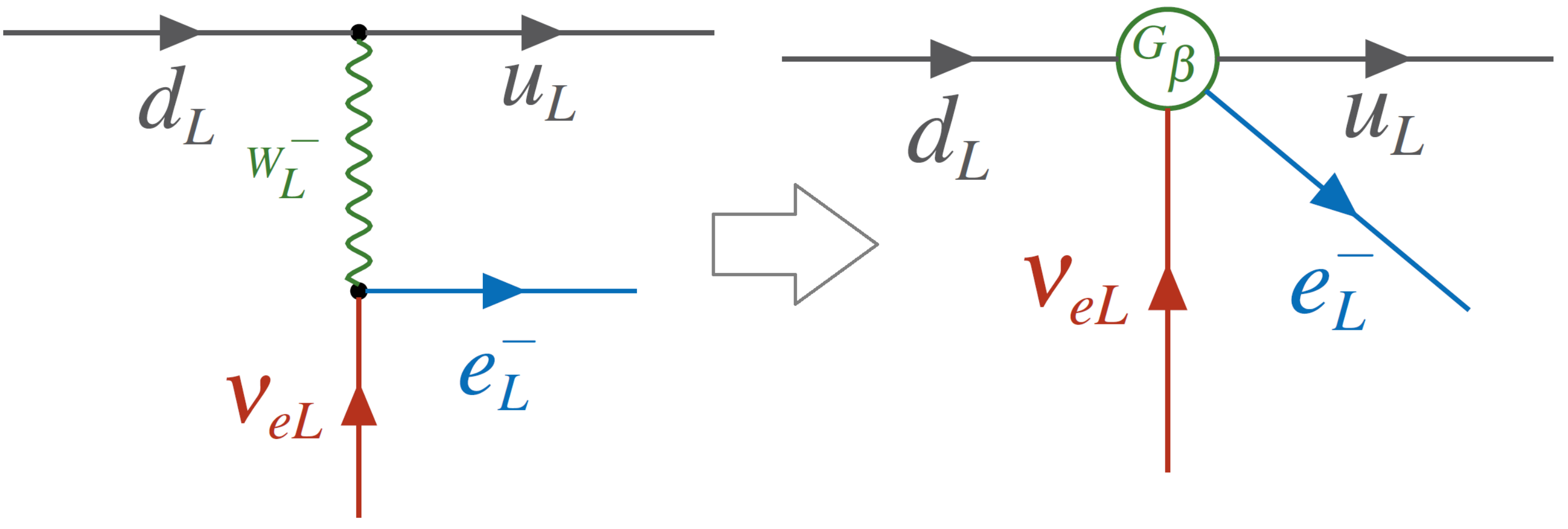}\label{Beta-GF}}
\qquad\subfloat[$W_R$-$W_L$ mediation for RH-LH mixed fields.]{\includegraphics[scale=0.27]{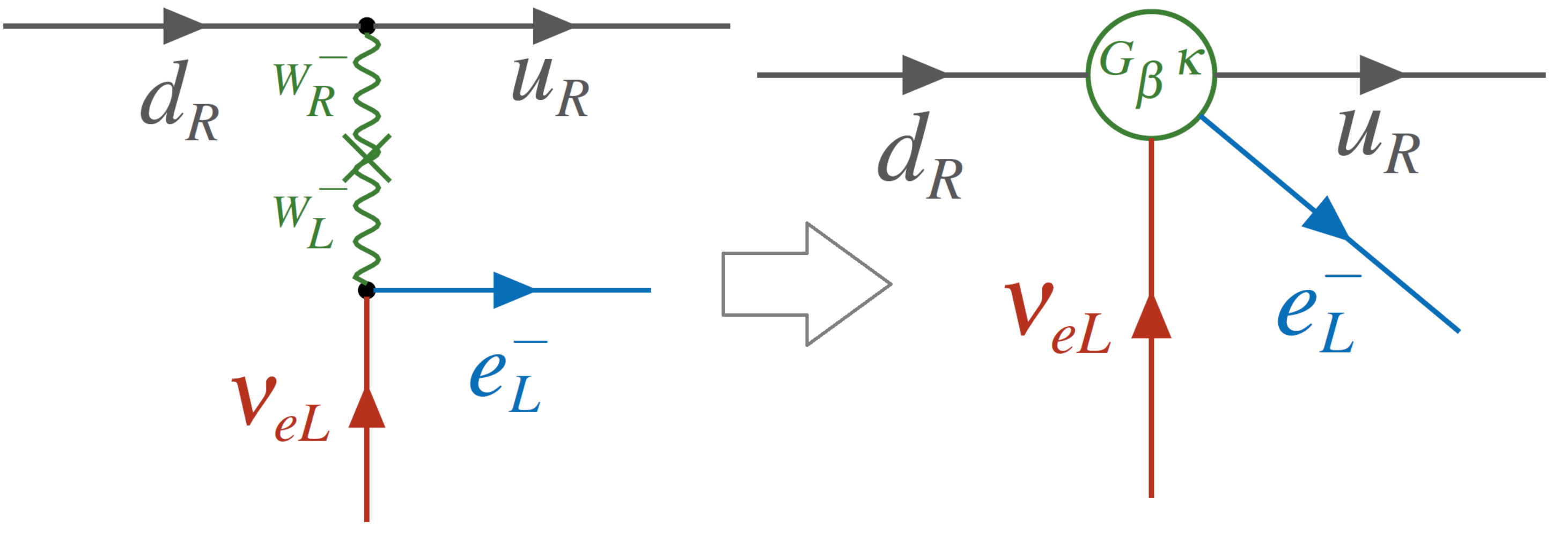}\label{Beta-Kappa}}

\subfloat[$W_L$-$W_R$ mediation for LH-RH mixed fields.]{\includegraphics[scale=0.27]{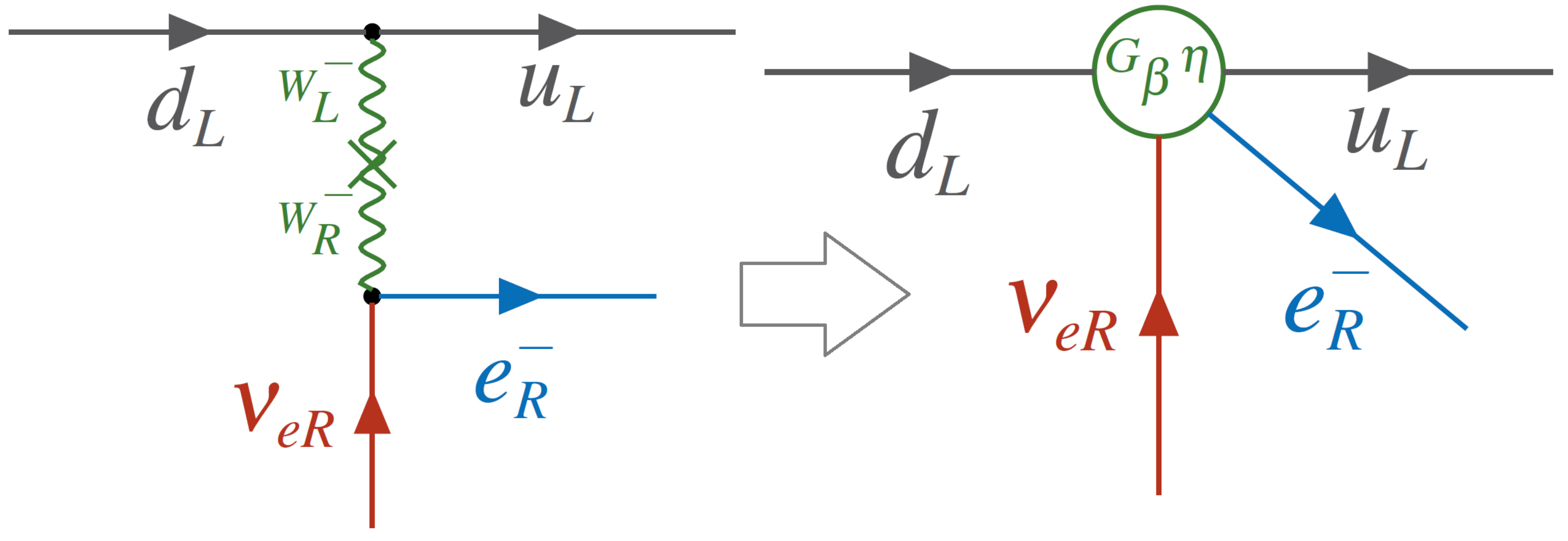}\label{Beta-Eta}}
\qquad\subfloat[$W_R$ mediation for purely RH fields.]{\includegraphics[scale=0.27]{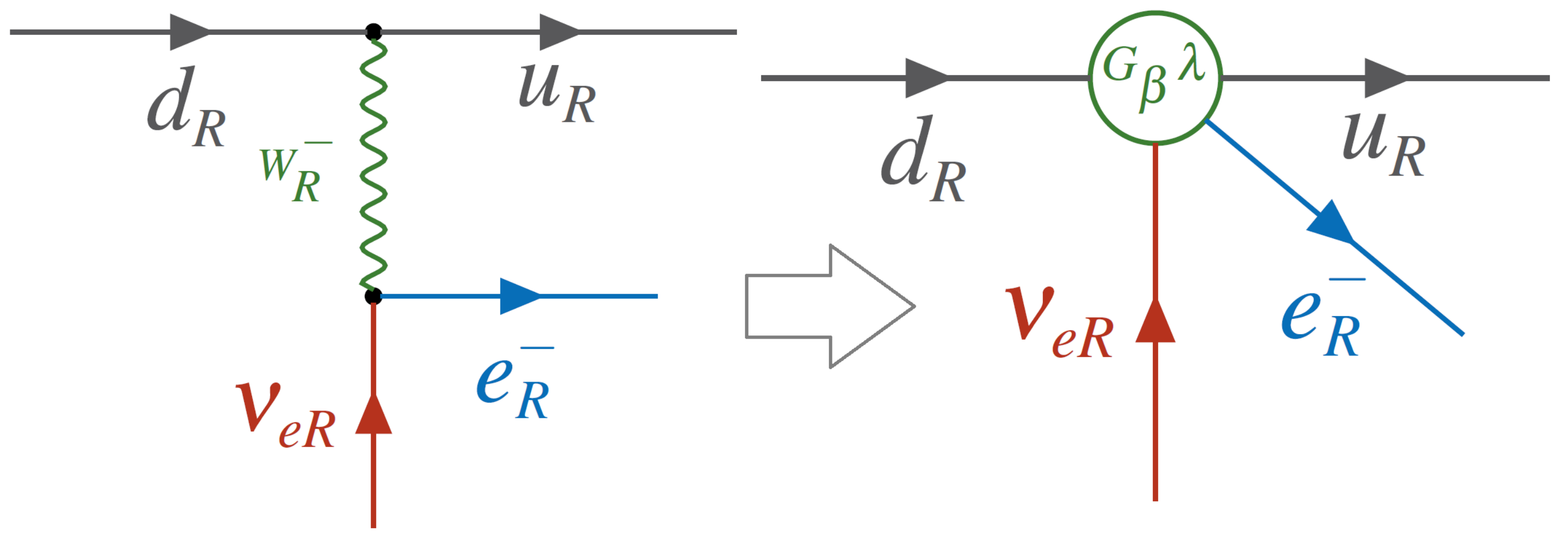}\label{Beta-Lambda}}
\caption{$\beta$-decay diagrams in LRSM at the $W$-boson and effective Fermi-like four-fermion level.\label{Beta-Dia}}
\end{figure*}
Starting from the charge-current Lagrangian of \eq{Weak-L} for the LRSM, after applying second-order perturbation in the gauge coupling $g$, we get four different types of $\beta$-decay diagrams due to the presence of RH currents (see \fg{Beta-Dia}). We can then integrate out the heavy degrees of freedom for the charged bosons ($m_{W_L}, m_{W_R}\geq 80$ GeV) to get point-like Fermi vertices. \fg{Beta-GF} shows the usual $\beta$-decay via $W_L^-$ exchange with $G_{\beta}=G_F\cos{\theta_c}$ being the effective point-like coupling between LH-quarks and LH-lepton currents, and $G_F$ is the Fermi constant. Figures. \ref{Beta-Kappa}, \ref{Beta-Lambda}, \ref{Beta-Eta} describe the presence of RH quarks and/or lepton currents. In \fg{Beta-Kappa} the RH-quark and LH-lepton currents are coupled by $W_R$-$W_L$ mixing, mediated by the effective coupling $G_{\beta}\kappa$. \fg{Beta-Eta} shows the diagram of $W_L$-$W_R$ exchange between LH-quarks and RH-lepton currents with effective coupling $G_{\beta}\eta$. Lastly, \fg{Beta-Lambda} shows the RH counterpart for the usual $\beta$-decay of \fg{Beta-GF} with $W_R^-$ exchange, and $G_{\beta}\lambda$ is the effective coupling between RH currents for quarks and lepton. The exact expressions for the effective couplings, $G_F$, $\lambda$, $\eta$, in terms of the LRSM parameters are given in Eqs. (7)-(9) of \rf{Hirsch:1995rf}. For small $W_L$-$W_R$ mixing ($\xi \ll 1$) we get,
\begin{align}
& G_F \simeq \sqrt{2}g^2/8m^2_{W_L}\quad ,\quad \eta=\kappa\simeq\tan{\xi},\\ 
&\lambda\simeq(m_{W_1}/m_{W_2})^2\simeq(m_{W_L}/m_{W_R})^2.
\end{align}
Thus at the level of effective couplings we can write an effective low-energy ($V\pm A$) Fermi-like current-current Lagrangian for $\beta$ decay\cite {Hirsch:1995rf, Doi+Kotani1985} considering the RH-currents. Taking $\cos{\theta^{\prime}_c}/\cos{\theta_c}=1$ one gets,
\begin{align}\label{Lbeta}
\mathcal{L}_{\text{\tiny{LR}}}^{\beta} = \frac{G_{\beta}}{\sqrt{2}} \[j^{\mu}_{L}J_{L\mu}^{\dagger} +\kappa j^{\mu}_{L}J_{R\mu}^{\dagger}+\eta j^{\mu}_{R}J_{L\mu}^{\dagger}&+\lambda j^{\mu}_{R}J_{R\mu}^{\dagger}\]+\text{H.c.},
\end{align}
where $j^{\mu}_{\alpha}=\overline{e_{\alpha}}\gm\nu_{e\alpha}$ and $J^{\dagger}_{\alpha,\mu}=\overline{u_{\alpha}}\gamma_{\mu} d_{\alpha}$ are leptonic and hadronic currents respectively with $\alpha=L,R$. The four terms in \eq{Lbeta}, in that order, correspond to the four diagrams of \fg{Beta-Dia}, respectively. H.c. denotes the Hermitian conjugate terms, which do not contribute to $\NDBD$.  

Notice that the neutrino fields are written in the flavor basis. The light and heavy neutrino mixing parameters in \eq{neu-mix}, are part of the leptonic currents and not of the effective BSM parameters $\eta$, $\lambda$. The LNV parameters of neutrino mixing are realized at the amplitude level in our analysis. \footnote{See Sec. \ref{EFT} for the effective-field theory approach to $\NDBD$ where the LNV parameters are interpreted at the effective coupling level but give us the same formula for the half-life.}

%\vspace{-5mm}
\subsection{Amplitudes and diagrams for $\NDBD$ from $\mathcal{L}_{\text{\tiny{LR}}}^{\beta}$.}
At the effective Lagrangian level of \eq{Lbeta} $\NDBD$ amplitude arises at second-order ($G^2_{\beta}$) of perturbation. The time-ordered product of $\mathcal{L}_{\text{\tiny{LR}}}^{\beta}$ has ten distinct terms,      
%\vspace{-20pt}
\begin{widetext}
\vspace{-5mm}
\begin{align}\label{TLL1}
&\mathcal{T}\(\mathcal{L}_{\text{\tiny{LR}}}^{\beta}(x)\mathcal{L}_{\text{\tiny{LR}}}^{\beta}(y)\)=\frac{G^2_{\beta}}{2}\mathcal{T}\([j_{L}J^{\dagger}_{L}]_{x}[j_{L}J^{\dagger}_{L}]_{y}+2\kappa[j_{L}J^{\dagger}_{L}]_{x}[j_{L}J^{\dagger}_{R}]_{y}+\kappa^2[j_{L}J^{\dagger}_{R}]_{x}[j_{L}J^{\dagger}_{R}]_{y}\non 
&+\lambda^2[j_{R}J^{\dagger}_{R}]_{x}[j_{R}J^{\dagger}_{R}]_{y}+2\lambda\eta[j_{R}J^{\dagger}_{R}]_{x}[j_{R}J^{\dagger}_{L}]_{y}+\eta^2[j_{R}J^{\dagger}_{L}]_{x}[j_{R}J^{\dagger}_{L}]_{y}\non 
&+\lambda[j_{L}J^{\dagger}_{L}]_{x}[j_{R}J^{\dagger}_{R}]_{y}+\eta[j_{L}J^{\dagger}_{L}]_{x}[j_{R}J^{\dagger}_{L}]_{y}+\kappa\lambda[j_{L}J^{\dagger}_{R}]_{x}[j_{R}J^{\dagger}_{R}]_{y}+\kappa\eta[j_{L}J^{\dagger}_{R}]_{x}[j_{R}J^{\dagger}_{L}]_{y}\).
\end{align}
\end{widetext}
From the above time-ordered product we see three types of combinations of leptonic currents: $j_Lj_L$, $j_Rj_R$, and $j_Lj_R$. After applying Wick's theorem to the time-ordered product, the neutrino fields in the leptonic currents get contracted, giving rise to the virtual neutrino propagator of $\NDBD$. The flavor neutrinos are linear combinations of mass eigenstates as in \eq{neu-mix}. Thus the virtual neutrino propagators would be of two types: light or heavy massive Majorana neutrinos \cite{Borah:2017ldt} for each of the three leptonic current combinations. Expressed in terms of the usual Dirac propagator we get for the neutrino propagators \cite{Srednicki2007},
\vspace{-3.5mm} 
\begin{align}
&\mathcal{T}\(j_L(x)j_L(y)\) \propto \contraction{}{\nu}{{}_{eL}(x)}{\nu}\nu_{eL}(x)\nu^{T}_{eL}(y)\non
&=\sum_i P_L \[U^2_{ei}\,\,S^{D}_{m_i}(x-y)+S^2_{ei}\,\,S^{D}_{M_i}(x-y)\]P_L C, \\ 
&\mathcal{T}\(j_R(x)j_R(y)\) \propto \contraction{}{\nu}{{}_{eR}(x)}{\nu}\nu_{eR}(x)\nu^{T}_{eR}(y)\non
&=\sum_i P_R\[T^{*2}_{ei}S^{D}_{m_i}(x-y)+V^{*2}_{ei}S^{D}_{M_i}(x-y)\]P_R C, \\
&\mathcal{T}\(j_L(x)j_R(y)\) \propto \contraction{}{\nu}{{}_{eL}(x)}{\nu}\nu_{eL}(x)\nu^{T}_{eR}(y)\non 
&=\sum_i P_L\[U_{ei}T^*_{ei}S^{D}_{m_i}(x-y)+S_{ei}V^*_{ei}S^{D}_{M_i}(x-y)\]P_R C,
\end{align}
where $C$ is the charge-conjugation matrix and the Dirac propagator $S^{D}_{m^{\prime}_i}$ is defined as ($m_i^{\prime}=m_i, M_i$),  
\begin{align}
S^{D}_{m^{\prime}_i}(x-y)=i\int\frac{d^4q}{(2\pi)^4}\frac{e^{-iq\cdot(x-y)}}{q^2-m^{\prime 2}_i}(\slashed{q}+m_i^{\prime}).
\end{align}
Because of the presence of the chiral projection operators $P_{L(R)}$ we will have two categories of contributions to the amplitude,  
\begin{align}
&\text{i)}\quad P_{L(R)} \frac{\slashed{q}+m^{\prime}_{i}}{q^{2}-m_{i}^{\prime 2}} P_{L(R)} \propto \frac{m^{\prime}_{i}}{q^{2}-m_{i}^{\prime 2}},\\
&\text{ii)}\quad P_{L(R)} \frac{\slashed{q}+m^{\prime}_{i}}{q^{2}-m_{i}^{\prime 2}} P_{R(L)} \propto \frac{\slashed{q}}{q^{2}-m_{i}^{\prime 2}}.
\end{align}
Thus we have (i) mass-dependent amplitudes where the two electrons have the same chirality, and (ii) momentum-dependent amplitudes when the two electrons have opposite chiralities \cite{PAS1999194}. The typical scale of momentum transfer for the vitual neutrino is $\abs{q}\simeq 100$ MeV. Here we assume $m_i\ll \abs{q}$ and $M_i\gg \abs{q}$ for the light and heavy Majorana neutrinos, respectively. Depending on the mass of the intermediate Majorana neutrinos, we have two categories of approximations for both the mass and momentum dependent amplitudes,  
\begin{align}
&\text{i) mass-dependent propagators:}\non &\frac{m^{\prime}_{i}}{q^{2}-m_{i}^{\prime 2}} \simeq\left\{\begin{array}{ll}{\frac{m_i}{q^{2}},} & {m_i^2 \ll q^{2}\qquad\text{light-}\nu_i} \\ {-\frac{1}{M_i},} & {M_i^2 \gg q^{2} \qquad\text{heavy-}N_i}\end{array}\right. \label{mass-app},\\ 
&\text{ii) momentum-dependent propagators:}\non
&\frac{\slashed{q}}{q^{2}-m_{i}^{\prime 2}} \simeq \left\{\begin{array}{cc}{\frac{1}{|q|},} & {m_i^2 \ll q^{2}\qquad\text{light-}\nu_i} \\ {-\frac{|q|}{M_i^2},} & {M_i^2 \gg q^{2}\qquad\text{heavy-}N_i}\end{array}\right. \label{mom-app} .
\end{align}
For the keV scale ($M_i<\abs{q}$) neutrino case see Refs. \cite{Barry:2014ika, Borah:2017ldt}. We now discuss the (i) mass-dependent and (ii) momentum-dependent cases separately.
\begin{figure*}[!]
\centering
\subfloat[Light neutrino exchange for purely LH currents. Diagram $\propto \eta_{m}$ arising from $j_{L}J_{L}^{\dagger}j_{L}J_{L}^{\dagger}$ term.\label{Left-Light}]{\includegraphics[scale=1.7]{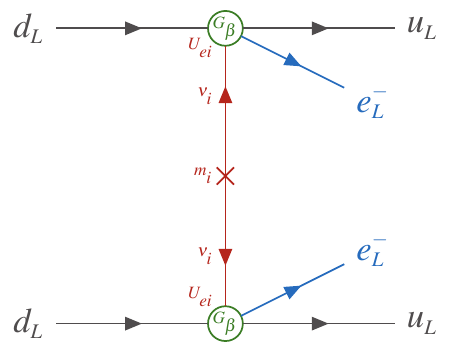}}\qquad
\subfloat[Heavy neutrino exchange for purely RH currents. Diagram $\propto \eta_{N}$ arising from $j_{R}J_{R}^{\dagger}j_{R}J_{R}^{\dagger}$ term.\label{Right-Heavy}]{\includegraphics[scale=1.7]{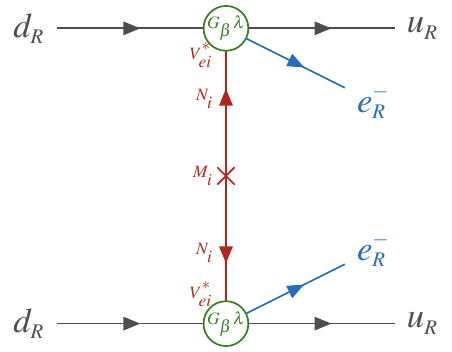}}
\caption{Relevant diagrams for $\NDBD$ in LRSM for both electron of same chirality.}\label{dia-same}
\end{figure*}
%\vspace{-5mm} 
\subsubsection*{Mass-dependent mechanisms: Outgoing electrons having same chirality}
The first six terms on the right-hand side of \eq{TLL1} are mass-dependent terms where both the electrons are either LH or RH. We can ignore most of the second-order terms because of the smallness of the BSM parameters ($\kappa$, $\lambda$, $\eta$ $\ll$ 1) for both light and heavy neutrino exchange. Moreover, the first-order term in $\kappa$ is further suppressed because of its dependence on neutrino mass. Since the mixing matrix $S$ is small and given the heavy mass $M_i$ being in the denominator, the heavy neutrino exchange case for purely LH currents can be ignored. Thus, the first term $[j_{L}J^{\dagger}_{L}]_{x}[j_{L}J^{\dagger}_{L}]_{y}$ gives rise to the regular mass mechanism of \fg{Left-Light} of light neutrino exchange for purely LH hadronic and leptonic currents. The amplitude for the `mass-mechanism' is then,  
\begin{align}
\mathcal{A}_L^{\nu} \propto G_{\beta}^{2} \sum_{i}\frac{U_{e i}^{2} m_{i}}{q^{2}},
\end{align}
where the dimensionless LNV complex parameter $\eta_m=\abs{\eta_{m}}\exp(i\phi_m)$ for the `mass-mechanism' along with the phase are defined in terms of the BSM parameters of LRSM as follows:
\begin{align}
\abs{\eta_{m}}&=\frac{1}{m_e}\abs{\effmass}=\frac{1}{m_e}\abs{\sum_i U^2_{ei}m_i},\label{eta_m}\\ 
\phi_m &=\text{Arg}\[\sum_i U^2_{ei}m_i\]. \label{phi_m}
\end{align}
The only second-order term considered in \eq{TLL1} is the $\lambda^2$ term for the heavy neutrino exchange because the mixing matrix $V$ is assumed to be large. Thus, from the term $\lambda^2[j_{R}J^{\dagger}_{R}]_{x}[j_{R}J^{\dagger}_{R}]_{y}$ we get the diagram of \fg{Right-Heavy}. Then, the amplitude for the heavy neutrino exchange for the purely RH currents is, 
\begin{align}
\mathcal{A}_R^{N} \propto G_{\beta}^{2}\lambda^2\sum_{i}\frac{V_{e i}^{* 2}}{M_{i}},
\end{align}
where the dimensionless LNV parameter $\eta_N=\abs{\eta_{N}}\exp(i\phi_N)$ for the heavy neutrino exchange ($N_i$) is,
\begin{align}
\abs{\eta_{N}} & =m_p\lambda^2\abs{\sum_i\frac{V^{*2}_{ei}}{M_i}}=m_p\(\frac{m_{W_L}}{m_{W_R}}\)^4\abs{\sum_i\frac{V^{*2}_{ei}}{M_i}},\label{eta_N}\\ 
\phi_N &=\text{Arg}\[\sum_i\frac{V^{*2}_{ei}}{M_i}\]. \label{phi_N}
\end{align}
%\vspace{-10mm}  
\subsubsection*{Momentum-dependent mechanisms: Outgoing electrons having opposite chiralities}
The last four terms in \eq{TLL1} are momentum-dependent terms. The first-order terms $\lambda$ and $\eta$ can give competing contributions to $\NDBD$ compared with the regular mass-mechanism of \fg{Left-Light} for light neutrino exchange. Thus, the term $\lambda[j_{L}J^{\dagger}_{L}]_{x}[j_{R}J^{\dagger}_{R}]_{y}$ gives rise to the diagram of \fg{Lambda-Light}, the so-called $\lambda$ mechanism, due to the combination of LH and RH currents. The amplitude of \fg{Lambda-Light} for the $\lambda$-mechanism is then,
\begin{figure*}[!]
\centering
\subfloat[$\lambda$-diagram due to both LH and RH currents. Diagram $\propto \eta_{\lambda}$ arising from $j_{L}J_{L}^{\dagger}j_{R}J_{R}^{\dagger}$ term.\label{Lambda-Light}]{\includegraphics[scale=1.7]{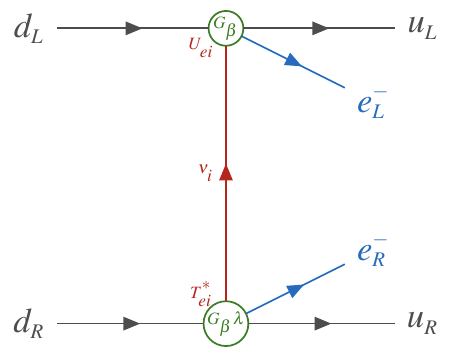}}\qquad
\subfloat[$\eta$-diagram due to gauge boson mixing. Diagram $\propto \eta_{\eta}$ arising from $j_{L}J_{L}^{\dagger}j_{R}J_{L}^{\dagger}$ term.\label{Eta-Light}]{\includegraphics[scale=1.7]{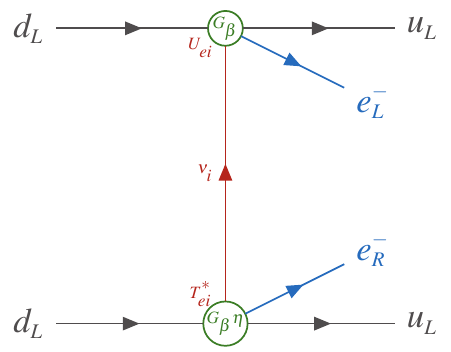}}
\caption{Relevant diagrams for $\NDBD$ in LRSM for both electrons of opposite chirality.}\label{dia-opp}
\end{figure*}
\begin{equation}
\mathcal{A}_{\lambda}^\nu \propto G_{\beta}^{2}\lambda \sum_{i} U_{e i} T_{e i}^{*} \frac{1}{q},
\end{equation}
where the corresponding dimensionless LNV PPP, $\eta_{\lambda}=\abs{\eta_{\lambda}}\exp(i\phi_{\lambda})$, 
\begin{align}
\abs{\eta_{\lambda}}&=\lambda\abs{\sum_i U_{ei}T^*_{ei}}=\(\frac{m_{W_L}}{m_{W_R}}\)^2\abs{\sum_i U_{ei}T^*_{ei}}\label{eta_lam},\\
\phi_\lambda &=\text{Arg}\[\sum_i U_{ei}T^*_{ei}\].\label{phi_lam}
\end{align}
The other first-order term $\eta[j_{L}J^{\dagger}_{L}]_{x}[j_{R}J^{\dagger}_{L}]_{y}$ in \eq{TLL1} gives rise to the diagram of \fg{Eta-Light}, the so-called $\eta$ mechanism due to $W_L-W_R$ mixing. The amplitude for \fg{Eta-Light} for the $\eta$ mechanism is then
\begin{equation}
\mathcal{A}_{\eta}^\nu \propto G_{\beta}^{2}\eta \sum_{i} U_{e i} T_{e i}^{*} \frac{1}{q},
\end{equation}
where the corresponding dimensionless LNV PPP, $\eta_{\eta}=\abs{\eta_{\eta}}\exp(i\phi_{\eta})$, is 
\begin{align}
\abs{\eta_{\eta}}&=\eta\abs{\sum_i U_{ei}T^*_{ei}}=\tan{\xi}\abs{\sum_i U_{ei}T^*_{ei}},\label{eta_eta}\\
\phi_\eta & =\text{Arg}\[\sum_i U_{ei}T^*_{ei}\].\label{phi_eta}
\end{align}
Terms due to heavy neutrino exchange are suppressed, being proportional to $S_{e i} V_{e i}^{*} q/M_{i}^{2}$ \cite{Barry:2013xxa}. 

Apart from the diagrams considered in Figs. {dia-same} and \ref{dia-opp}, there could be additional contributions due to exchange of $SU(2)_R$ and $SU(2)_L$ Higgs triplets in LRSM, see Fig. 3 of \rf{Barry:2013xxa}. These diagrams are suppressed \cite{Barry:2013xxa, Dev:2014xea, Tello:2010am} and hence we will not consider them in the subsequent analysis.         
\subsection{Half-Life for $\NDBD$}
Considering the total amplitude for $\NDBD$ for the four diagrams of \fg{dia-same} and \ref{dia-opp}, 
\begin{align}
\mathcal{A}^{0\nu}=\mathcal{A}_{L}^{\nu}+\mathcal{A}_{R}^{N}+\mathcal{A}_{\lambda}^\nu+\mathcal{A}_{\eta}^\nu
\end{align}
we arrive at the following inverse half-life formula for $\NDBD$,
\begin{align}\label{half-life1}
[T^{0\nu}_{1/2}]^{-1}=g^4_A &\[C_{m}\abs{\eta_{m}}^2+C_{N}\abs{\eta_{N}}^2+C_{\lambda}\abs{\eta_{\lambda}}^2+C_{\eta}\abs{\eta_{\eta}}^2\non &+\sum_{i\neq j}^{\{m,N,\lambda,\eta\}}C_{ij}\abs{\eta_i}\abs{\eta_j}\cos{(\phi_i-\phi_j)}\],
\end{align}
where we have factorized $g_A^4=(1.27)^4$ to be consistent with our definitions of the PSFs \cite{PhysRevC.88.037303, Neacsu:2015uja}, see below. The first four terms are contributions of the individual mechanisms. The rest of the terms are due to the interference between pairs of mechanisms, we have six such combinations. The differences in phases for the LNV parameters $\eta_i$s (Eqs. (\ref{eta_m}), (\ref{eta_N}), (\ref{eta_lam}), (\ref{eta_eta})) may produce interference effects. The $C_i$ and $C_{ij}$ are products of relevant NME and PSF for individual and interference terms, respectively \cite{Doi+Kotani1985, SuhonenCivitarese1998}: 
\begin{align}
&C_m =G_{01}\[M_{GT}-\(\frac{g_V}{g_A}\)^2 M_F+M_T\]^2, \label{C_m}\\
&C_N=G_{01}\[M_{GTN}-\(\frac{g_V}{g_A}\)^2 M_{FN}+M_{TN}\]^2, \label{C_N}\\
&C_\lambda =G_{02}\mathcal{M}^2_{2-}-\frac{2}{9}G_{03}\mathcal{M}_{1+}\mathcal{M}_{2-}+\frac{1}{9}G_{04}\mathcal{M}_{1+}^2\label{C_lambda}\\
&C_\eta =G_{02}\mathcal{M}^2_{2+}-\frac{2}{9}G_{03}\mathcal{M}_{1-}\mathcal{M}_{2+}+\frac{1}{9}G_{04}\mathcal{M}^2_{1-}\non &\qquad\qquad-G_{07}M_P M_R+G_{08}M_P^2+G_{09}M_R^2,\label{C_eta}\\
&C_{mN} =-2G^{\prime}_{01}\[M_{GT}-\(\frac{g_V}{g_A}\)^2 M_F+M_T\]\non
&\qquad\qquad\qquad\times\[M_{GTN}-\(\frac{g_V}{g_A}\)^2 M_{FN}+M_{TN}\],\label{C_m-N}\\
&C_{m\lambda}=-\[M_{GT}-\(\frac{g_V}{g_A}\)^2 M_F+M_T\]\non
&\qquad\qquad\times\[G_{03}\mathcal{M}_{2-}-G_{04}\mathcal{M}_{1+}\],\label{C_m-lambda}\\
&C_{N\lambda}=-\[M_{GTN}-\(\frac{g_V}{g_A}\)^2 M_{FN}+M_{TN}\]\non
&\qquad\qquad\times\[G_{03}\mathcal{M}_{2-}-G_{04}\mathcal{M}_{1+}\],\label{C_N-lambda}\\
&C_{m\eta}=\[M_{GT}-\(\frac{g_V}{g_A}\)^2 M_F+M_T\]\non &\,\,\,\,\,\,\,\times\[G_{03}\mathcal{M}_{2+}-G_{04}\mathcal{M}_{1-}-G_{05}M_P+G_{06}M_R\],\label{C_m-eta}\\
&C_{N\eta}=\[M_{GTN}-\(\frac{g_V}{g_A}\)^2 M_{FN}+M_{TN}\]\non &\,\,\,\,\,\,\,\times\[G_{03}\mathcal{M}_{2+}-G_{04}\mathcal{M}_{1-}-G_{05}M_P+G_{06}M_R\],\label{C_N-eta}\\
&C_{\lambda\eta}=-2G_{02}\mathcal{M}_{2-}\mathcal{M}_{2+}+\frac{2}{9}G_{03}\[ \mathcal{M}_{1+}\mathcal{M}_{2+}\non & \qquad\qquad\qquad +\mathcal{M}_{2-}\mathcal{M}_{1-}\]-\frac{2}{9}G_{04}\mathcal{M}_{1+}\mathcal{M}_{1-},\label{C_lambda-eta}
\end{align}
where the following definitions are used,
\begin{align}
&\mathcal{M}_{1\pm}=M_{GTq}\pm 3\(\frac{g_V}{g_A}\)^2 M_{Fq}-6M_{Tq}, \\
&\mathcal{M}_{2\pm}=M_{GT\omega}\pm \(\frac{g_V}{g_A}\)^2 M_{F\omega}-\frac{1}{9}\mathcal{M}_{1\mp}. \label{M2pm}
\end{align}
Note that the term $\frac{1}{9}\mathcal{M}_{1\mp}$ in \eq{M2pm} above is the correct expression (see footnote on p.146 of \rf{10.1143/ptp/89.1.139}); it was incorrectly written as $\frac{1}{9}\mathcal{M}_{1\pm}$ in Eq. (3.5.16) of \rf{Doi+Kotani1985}. Detailed expressions for the thirteen NME $\{M_F$, $M_{GT}$, $M_T$, $M_{F\omega}$, $M_{Fq}$, $M_{GT\omega}$, $M_{GTq}$, $M_{Tq}$, $M_P$, $M_R$, $M_{FN}$, $M_{GTN}\}$, $M_{TN}\}$  are given in the appendix of \rf{PhysRevC.98.035502}. The expressions for the nine PSF integrals $\{G_{01}-G_{09}\}$ are \cite{Neacsu:2015uja}
\begin{align}\label{PSF}
G_{0k}=\frac{g^{0\nu}}{r_A^2}\int_1^{T+1}\!\!\!\!\!\!\!\! b_{0k}F_0(Z_s, \epsilon_1)F_0(Z_s, \epsilon_2)p_1p_2\epsilon_1\epsilon_2\text{d}\epsilon_1,
\end{align}
with
\begin{align}
g^{0\nu}=\frac{(G_F \cos{\theta_c})^4 m_e^9}{(2\pi)^5 \ln 2}=2.8\times 10^{-22}\quad\text{yr}^{-1},
\end{align}
where the expressions for the nine kinematical factors $b_{0k}$ ($k=1\sim 9$) and definitions of other terms are given in Appendix A of \rf{Neacsu:2015uja}. The PSF $G^{\prime}_{01}$ in $C_{mN}$ (\eq{C_m-N}) for the interference between regular `mass-mechanism' (\fg{Left-Light}) and heavy-neutrino exchange for purely RH currents (\fg{Right-Heavy}) has the same expression as $G_{01}$ of \eq{PSF} without the factors $\epsilon_1\epsilon_2$ \cite{Ahmed:2017pqa}.  Because of our definitions of the PSFs and NMEs, the products $C_i$s and $C_{ij}$s are reported in the units of $\text{y}^{-1}$.
\vspace{-3mm} 
\subsection{Effective-Field Theory approach to $\NDBD$}\label{EFT}
Before proceeding to the analysis section we would like to point out that the effective Lagrangian of \eq{Lbeta} arises from an explicit LRSM charge-current Lagrangian, \eq{Weak-L}. This is exactly the approach taken in the standard literature, e.g. as in Ref. \cite{Doi+Kotani1985}, where RH neutrinos are assumed to contribute besides the usual SM neutrinos. In the effective-field theory (EFT) approach to $\NDBD$ we encounter a dimension-six Lagrangian \cite{PhysRevC.98.035502, Deppisch2012} that is similar in structure to $\mathcal{L}_{\text{\tiny{LR}}}^{\beta}$ of \eq{Weak-L},
\begin{align}\label{EFT-L6}
\mathcal{L}^{\text{\tiny{EFT}}}_{6} = &\frac{G_{\beta}}{\sqrt{2}} \[j^{\mu}_{V-A}J_{V-A,\mu}^{\dagger}+\varepsilon_{V-A}^{V+A} j^{\mu}_{V+A}J_{V-A,\mu}^{\dagger}\non &+\varepsilon_{V+A}^{V+A} j^{\mu}_{V+A}J_{V+A,\mu}^{\dagger}+\varepsilon_{S-P}^{S+P}j_{S+P}J_{S-P}^{\dagger}\non &+\varepsilon_{S+P}^{S+P}j_{S+P}J_{S+P}^{\dagger}+\varepsilon_{T_{R}}^{T_{R}}j^{\mu\nu}_{T_{R}}J_{T_{R},\mu\nu}^{\dagger}\],
\end{align}
which is the most general Lorentz-invariant Lagrangian responsible for $\NDBD$ in the second order of perturbation theory. The leptonic and hadronic currents of the EFT Lagrangian are respectively $j_{\beta}=\overline{e}\mathcal{O}_{\beta}\nu$ and $J^{\dagger}_{\alpha}=\overline{u}\mathcal{O}_{\alpha}d$, with the $\mathcal{O}_{\alpha, \beta}$ operators defined as,
\begin{align}
\mathcal{O}_{V\pm A}=&\gamma^{\mu}\left(1\pm\gamma_{5}\right),\quad  \mathcal{O}_{S\pm P}=\left(1\pm\gamma_{5}\right),\non \mathcal{O}_{T_{R}}&=\frac{\mathrm{i}}{2}\left[\gamma_{\mu}, \gamma_{v}\right]\left(1+\gamma_{5}\right).
\end{align}
Note that the neutrino fields used in \eq{EFT-L6} are the SM LH-neutrinos in the flavor basis. Heavy RH neutrinos in \eq{neu-mix} are integrated out and any related parameters are absorbed in the definition of the effective BSM couplings $\varepsilon_{\alpha}^{\beta}$s. EFT formalism allows us to relate BSM physics parameters through the SM degrees of freedom. In the case of LRSM we approximate the effective BSM couplings as,
\begin{align}
\varepsilon_{V-A}^{V+A}=\eta_\eta \qquad , \qquad \varepsilon_{V+A}^{V+A}=\eta_\lambda
\end{align}
The scalar-pseudoscalar ($S\pm P$) and tensor ($T_R$) terms do not arise from the LRSM charged-current Lagrangian, but from other BSM models. The term related to the heavy-neutrino exchange in the presence of purely RH currents, $\mathcal{A}_{R}^{N}$ (\fg{Right-Heavy}), is not given by the $\mathcal{L}^{\text{\tiny{EFT}}}_{6}$ since it is a short-range contribution due to the exchange of heavy particles. $\mathcal{L}^{\text{\tiny{EFT}}}_{6}$ gives rise to long-range contributions to $\NDBD$ due to the exchange of light neutrinos, see Figs. (1(b)) and (1(c)) of \rf{PhysRevC.98.035502}. In the EFT approach to $\NDBD$ the dimension-nine Lagrangian is \cite{PhysRevC.98.035502}
\begin{align}\label{EFT-L9}
\mathcal{L}^{\text{\tiny{EFT}}}_{9} = &\frac{G^2_{\beta}}{2m_P} \[\epsilon_1 JJj+\epsilon_2 J^{\mu\nu}J_{\mu\nu}j+\epsilon_3^{LLz} J^{\mu}J_{\mu}j\non &+\epsilon_3^{RRz}J^{\mu}J_{\mu}j+\epsilon_3^{LRz} J^{\mu}J_{\mu}j+\epsilon_3^{RLz} J^{\mu}J_{\mu}j\non &+\epsilon_4 J^{\mu}J_{\mu\nu}j^{\nu}+\epsilon_5 J^{\mu}Jj_{\mu}\].
\end{align}
The expressions for the leptonic and hadronic currents are given in Ref. \cite{PhysRevC.98.035502}. The short-range contribution (see Fig. 1(d) of Ref. \cite{PhysRevC.98.035502}) to $\NDBD$, $\mathcal{A}_{R}^{N}$, arises from the $J^{\mu}J_{\mu}j$ term of $\mathcal{L}^{\text{\tiny{EFT}}}_{9}$ in first-order of perturbation where we approximate $\epsilon_3^{RRz}=\eta_N$. However, the $\NDBD$ half-life formula, \eq{half-life1}, is the same in both approaches. Thus, our analysis of the interference between different mechanisms arising from $\mathcal{L}_{\text{\tiny{LR}}}^{\beta}$ can easily be extended to a subset of terms of EFT approach to $\NDBD$ Lagrangians $\mathcal{L}^{\text{\tiny{EFT}}}_{6}$ and $\mathcal{L}^{\text{\tiny{EFT}}}_{9}$. For a complete discussion of $\NDBD$ in the EFT approach see Refs. \cite{Cirigliano:2018yza, delAguila:2012nu, Deppisch2012}. The contribution of the ($S\pm P$) and $T_R$ terms of $\mathcal{L}^{\text{\tiny{EFT}}}_{6}$ to the total decay rate of $\NDBD$, along with the constraints on the effective LNV couplings, have been studied with the assumption that the interference terms are negligible \cite{PhysRevC.98.035502}. As an extension of our current work, we plan to explore in the future the contribution of all the possible interference terms arising from $\mathcal{L}^{\text{\tiny{EFT}}}_{6}$. A similar analysis can be also carried out for the interference terms arising from $\mathcal{L}^{\text{\tiny{EFT}}}_{9}$; see, e.g., Eq. (5) of \cite{PhysRevC.98.035502}.       
%%%%%%%%%%%%%%%%%%%%%%%%%%%%%%%%%%%%%%%%%%%%%%%%%%%%%%%%%%%%%%%%%%%%%%%%%%%%%%%%%%%%%%%%%%%%%%%%%%%%

%%%%%%%%%%%%%%%%%%%%%%%%%%%%%%%%%%%%%%%%%%%%%%%%%%%%%%%%%%%%%%%%%%%%%%%%%%%%%%%%%%%%%%%%%%%%%%%%%%%%
\section{Analysis of interference terms}\label{Analysis}
\begingroup
\squeezetable
\begin{table}[!]
\centering
\caption{Values of the product of NME and PSF, $C_i$ and $C_{ij}$, for various nuclei for the $0^+\to 0^+$ transition in units of $\text{y}^{-1}$. See Tables \ref{PSF-values} and \ref{NME-values} of the appendix. \label{Ci-table}}
\begin{ruledtabular}
\begin{tabular}{lccccccc}

&               						&$^{48}$Ca 	&$^{76}$Ge 	&$^{82}$Se 	&$^{124}$Sn	&$^{130}$Te 	&$^{136}$Xe \\ 
\hline
&$C_m\cdot 10^{14}$ 	    				&$2.57$   		&$3.00$  		&$11.54$  		&$4.14$			&$5.22$  		&$4.39$  \\ 

&$C_N\cdot 10^{10}$ 	    				&$1.63$   		&$0.87$  		&$3.28$  		&$1.84$			&$2.25$  		&$1.86$  \\

&$C_{\lambda}\cdot 10^{13}$       	&$1.22$    		&$0.43$   		&$3.52$   		&$0.79$			&$1.24$   		&$0.99$  \\

&$C_{\eta}\cdot 10^{09}$        		&$1.45$   		&$1.40$  		&$5.11$  		&$2.74$			&$3.67$   		&$3.09$  \\

&$C_{mN}\cdot 10^{13}$      			&$-1.82$  		&$-4.11$  		&$-9.35$  		&$-6.10$			&$-6.64$  		&$-5.75$   \\

&$C_{m\lambda}\cdot 10^{14}$      	&$-0.90$  		&$-1.13$  		&$-5.68$  		&$1.97$			&$-2.64$  		&$-2.20$   \\

&$C_{m\eta}\cdot 10^{11}$       		&$0.38$   		&$0.64$   		&$1.91$   		&$-0.97$			&$1.19$   		&$1.01$  \\

&$C_{N\lambda}\cdot 10^{12}$      	&$-0.72$  		&$-0.61$  		&$-3.03$  		&$1.31$			&$-1.74$  		&$-1.43$   \\

&$C_{N\eta}\cdot 10^{10}$      		&$3.05$  		&$3.43$  		&$10.19$  		&$-6.45$			&$7.80$  		&$6.58$   \\

&$C_{\lambda\eta}\cdot 10^{13}$      &$-1.51$   		&$-0.60$  		&$-5.05$   		&$-1.06$			&$-1.65$   		&$-1.31$  \\

\end{tabular}
\end{ruledtabular}
\end{table}
\endgroup
\begin{table}[!]
\centering
\caption{Interference coefficients $\varepsilon_{m\lambda}(\alpha)$ in $\%$ for specific $\alpha$ values.} \label{ratios-table1}
\begin{ruledtabular}
\begin{tabular}{ccccc}
\multicolumn{1}{c}{Nuclei}    
& $\Large{\varepsilon_{m\lambda}}(0.25)$ 
& $\Large{\varepsilon_{m\lambda}}(0.5)$
& $\Large{\varepsilon_{m\lambda}}(0.75)$
& $\Large{\varepsilon_{m\lambda}}(1)$ \\
\colrule
$^{48}$Ca   &$6.42$ 		&$7.57$  	&$7.95$ 		&$8.03$ \\
                          							  
$^{76}$Ge   &$12.68$ 		&$14.94$ 	&$15.69$ 	&$15.85$  \\                                                                                                

$^{82}$Se   &$11.27$ 		&$13.28$ 	&$13.94$ 	&$14.08$  \\
                                            							   
$^{124}$Sn  &$13.81$ 		&$16.28$ 	&$17.09$ 	&$17.27$  \\

$^{130}$Te  &$13.16$ 		&$15.51$ 	&$16.28$ 	&$16.45$  \\

$^{136}$Xe  &$13.33$ 		&$15.70$  	&$16.49$  	&$16.66$  \\ 
\end{tabular}
\end{ruledtabular}
\end{table}
\begin{table}[!]
\centering
\caption{Interference coefficients $\varepsilon_{m\eta}(\alpha)$ in $\%$ for specific $\alpha$ values.} \label{ratios-table2}
\begin{ruledtabular}
\begin{tabular}{ccccc}
\multicolumn{1}{c}{Nuclei}    
& $\Large{\varepsilon_{m\eta}}(0.25)$ 
& $\Large{\varepsilon_{m\eta}}(0.5)$
& $\Large{\varepsilon_{m\eta}}(0.75)$
& $\Large{\varepsilon_{m\eta}}(1)$ \\
\colrule
$^{48}$Ca   &$25.11$  	&$29.60$   	&$31.07$  	&$31.40$   \\                          							  

$^{76}$Ge   &$39.27$ 		&$46.28$   	&$48.59$ 	&$49.09$   \\                                                                                                

$^{82}$Se   &$31.48$ 		&$37.10$   	&$38.94$ 	&$39.35$  \\                                            							   

$^{124}$Sn  &$36.32$ 		&$42.80$   	&$44.93$ 	&$45.40$  \\

$^{130}$Te  &$34.29$  	&$40.41$   	&$42.42$ 	&$42.86$  \\

$^{136}$Xe  &$34.75$ 		&$40.95$   	&$42.99$ 	&$43.44$  \\ 
\end{tabular}
\end{ruledtabular}
\end{table}
\begin{table}[!]
\centering
\caption{Interference coefficient $\varepsilon_{\lambda\eta}(\alpha)$ in $\%$ for specific $\alpha$ values.} \label{ratios-table3}
\begin{ruledtabular}
\begin{tabular}{ccccc}
\multicolumn{1}{c}{Nuclei}    
& $\Large{\varepsilon_{\lambda\eta}}(0.25)$ 
& $\Large{\varepsilon_{\lambda\eta}}(0.5)$
& $\Large{\varepsilon_{\lambda\eta}}(0.75)$
& $\Large{\varepsilon_{\lambda\eta}}(1)$ \\
\colrule

$^{48}$Ca   &$0.45$ 		&$0.54$  	&$0.56$ 		&$0.57$  \\
                          							  
$^{76}$Ge   &$0.31$ 		&$0.37$  	&$0.38$ 		&$0.39$  \\                                                                                                

$^{82}$Se   &$0.48$ 		&$0.56$  	&$0.59$ 		&$0.60$  \\ 
                                           							   
$^{124}$Sn  &$0.29$ 		&$0.34$  	&$0.35$ 		&$0.36$  \\

$^{130}$Te  &$0.31$ 		&$0.36$  	&$0.38$ 		&$0.39$  \\

$^{136}$Xe  &$0.30$ 		&$0.35$  	&$0.37$ 		&$0.38$  \\ 
\end{tabular}
\end{ruledtabular}
\end{table}
\begin{figure*}[!t]
\centering
\includegraphics[scale=0.6]{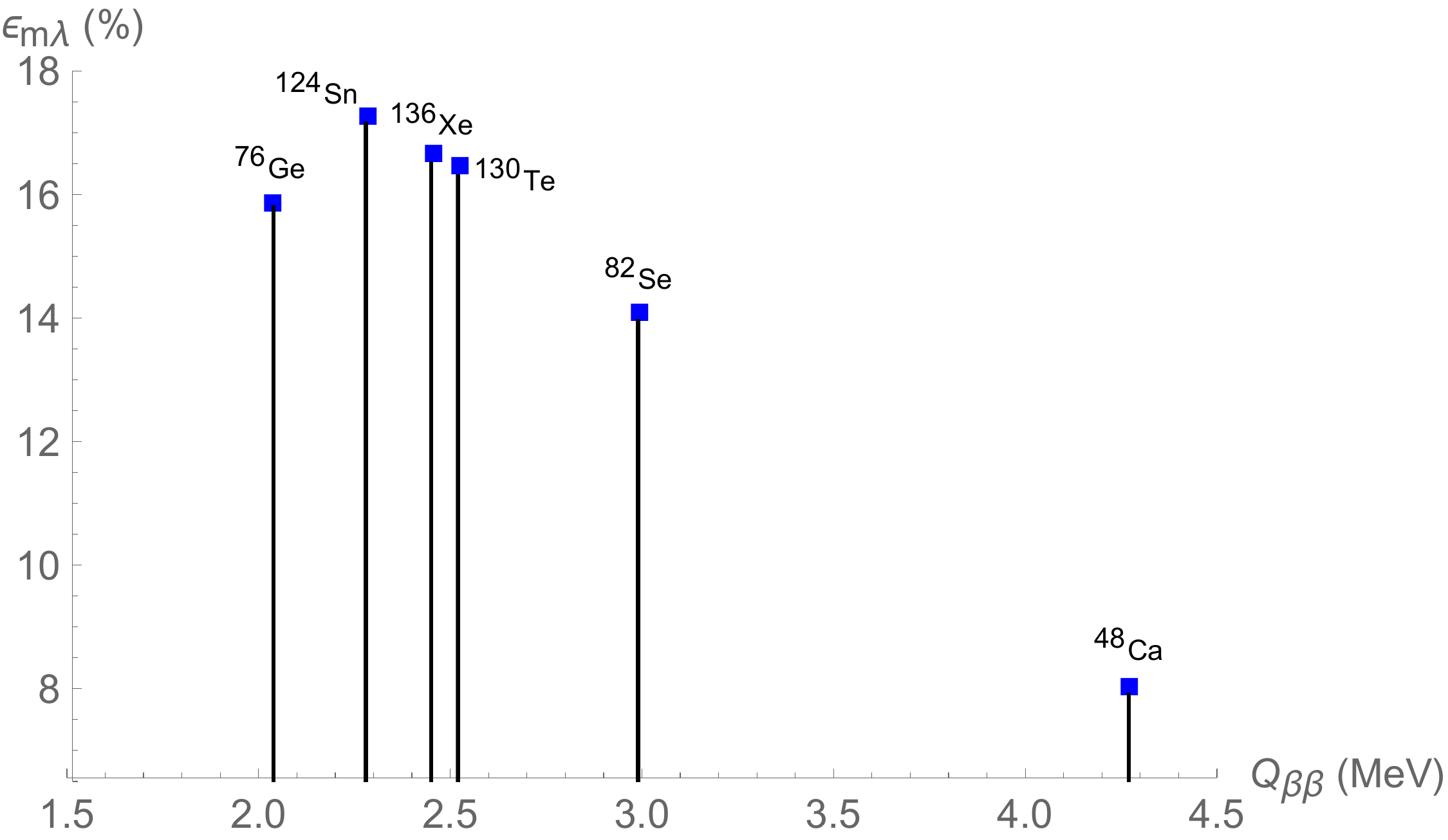}
\caption{Coefficient of maximum interference $\varepsilon_{m\lambda}(1)$ plotted against $Q_{\beta\beta}$ values. \label{m-lambda}}
\end{figure*}
\begin{figure*}[!t]
\centering
\includegraphics[scale=0.6]{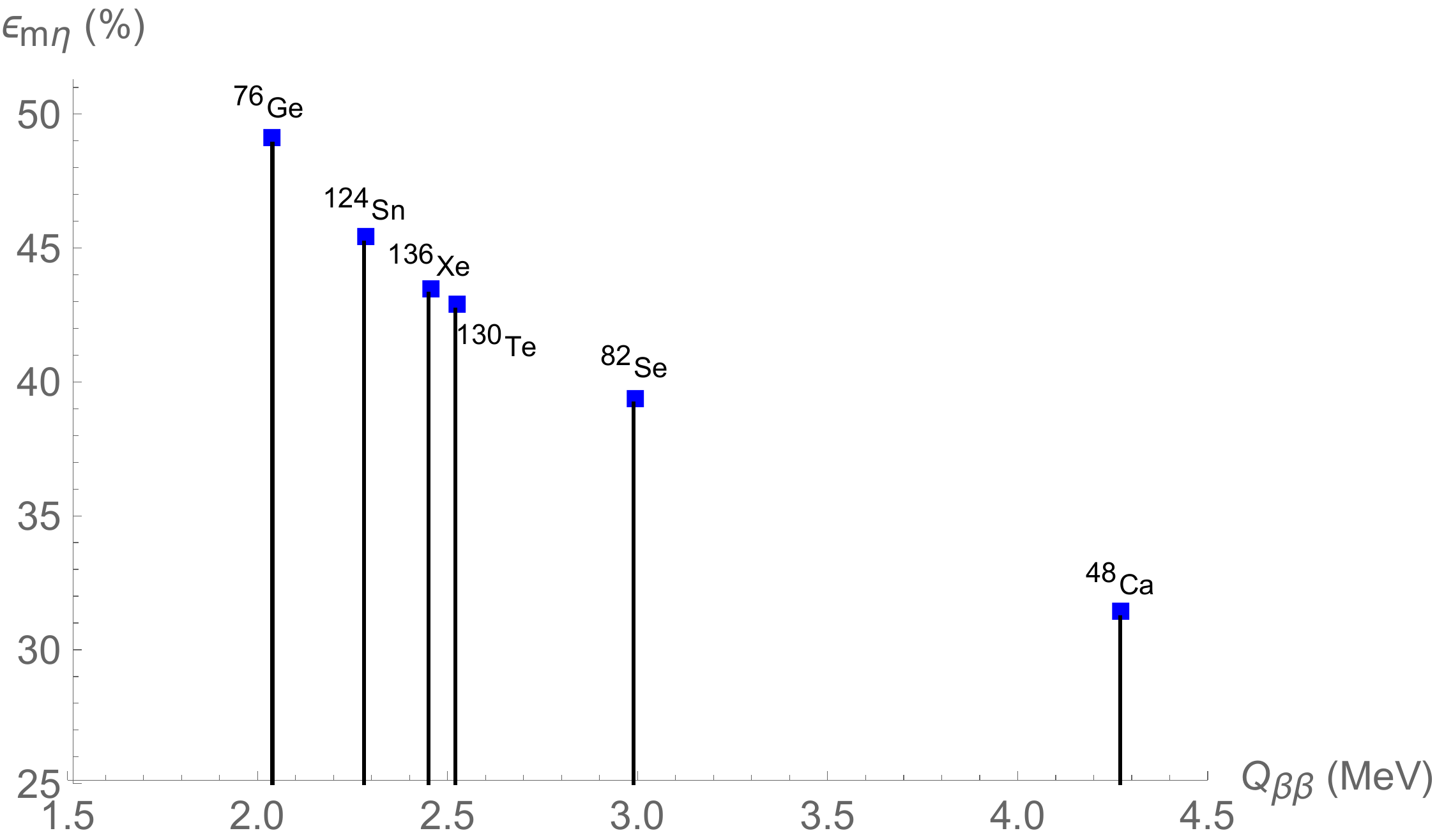}
\caption{Coefficient of maximum interference $\varepsilon_{m\eta}(1)$ plotted against $Q_{\beta\beta}$ values. \label{m-eta}}
\end{figure*}
\begin{figure*}[!t]
\centering
\includegraphics[scale=0.6]{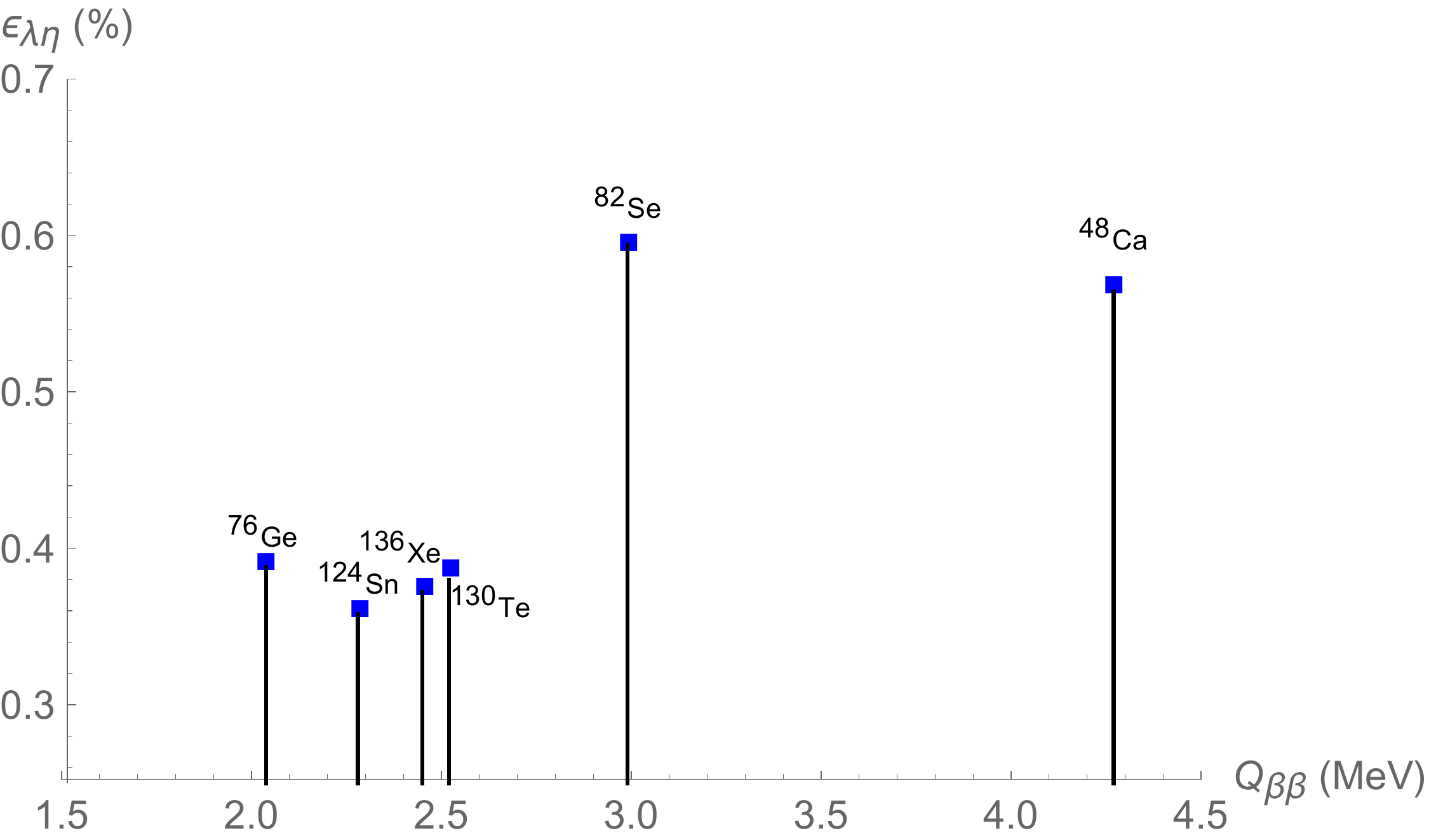}
\caption{Coefficient of maximum interference $\varepsilon_{\lambda\eta}(1)$ plotted against $Q_{\beta\beta}$ values. \label{lambda-eta}}
\end{figure*}
We now analyze the contribution of each of the interference terms in \eq{half-life1} by comparison with the related pairs of squared amplitudes for each individual mechanisms. The interference between light-LH and heavy-RH neutrinos ($C_{mN}$ term in \eq{half-life1}) was analyzed in \rf{Ahmed:2017pqa}. Here we analyze the other five terms (three after symmetry, see below). We write a generic approximate inverse half-life formula for a pair of mechanisms in the following manner:
\begin{align}\label{gen-half-life}
[T^{0\nu}_{1/2}]^{-1}\simeq g^4_A &\[C_i\abs{\eta_i}^2+C_j\abs{\eta_j}^2\non &+C_{ij}\abs{\eta_i}\abs{\eta_j}\cos{(\phi_i-\phi_j)}\],
\end{align}
where $i,\,j=\{m,\,N,\,\lambda,\,\eta\}$ and $i\neq j$. We assume the individual mechanism squared amplitude to be a factor $\alpha$ of each other ($0<\alpha\leq 1$),
\begin{align}\label{equality11}
&C_j\abs{\eta_j}^2=\alpha C_i\abs{\eta_i}^2 \Rightarrow\abs{\eta_j}=\sqrt{\alpha\frac{C_i}{C_j}}\abs{\eta_i}.
\end{align}
Thus, our approximate generic half-life expression becomes,
\begin{align}
[T^{0\nu}_{1/2}]^{-1}\simeq g^4_A(1+\alpha) C_i\abs{\eta_i}^2\[1+\varepsilon_{ij}\cos{(\phi_i-\phi_j)}\],
\end{align}
where the interference coefficient  
\begin{equation}\label{epsilon}
\varepsilon_{ij}(\alpha)=\frac{\sqrt{\alpha}}{1+\alpha}\frac{\abs{C_{ij}}}{\sqrt{\abs{C_i}\abs{C_j}}},
\end{equation}
would allow us to compare the contribution of the interference term with respect to that of each individual mechanisms for maximum interference, $|\cos{(\phi_i-\phi_j)}|=1$. We numerically calculate the products of NME and PSF, and the ten $C_i$ and $C_{ij}$ of \eq{C_m}-(\ref{C_lambda-eta}), given in Table \ref{Ci-table}. 

The NME for the six isotopes used in this study were calculated by shell-model techniques \cite{HoroiStoica2010,PhysRevC.87.014320} in three different model spaces, using three different effective Hamiltonians \cite{NeacsuHoroi2015,PhysRevC.93.024308,PhysRevC.98.035502}. Some of the NME are sensitive to short-range correlations (SRC) effects entering the two-body matrix elements. Here we used the CD-Bonn SRC parametrization \cite{PhysRevC.87.014320}. Using the AV18 SRC parametrization \cite{PhysRevC.87.014320}, or/and the Strasbourg-Madrid choice for the effective Hamiltonians \cite{PhysRevC.98.035502} does not significantly change the results. The relevant NME and PSF used in this study are given in the Appendix. Besides the values of \rf{PhysRevC.98.035502}, we have also considered the PSF of \rf{Stefanik:2015twa} in conjungtion to the various sets of NME. The results for the two sets of PSF do not have any appreciable difference. As discussed in Ref. \cite{PhysRevC.98.035502}, competing contributions to the NME are always present but some are dominant, such as those of the Gamow-Teller type operators, thus avoiding full cancellations of the total NME. In addition, given that calculations in different model spaces with different effective Hamiltonians lead to similar results, we have confidence in the reliability of our conclusions. 

Using \eq{epsilon} we then evaluate the interference coefficients, $\varepsilon_{m\lambda}$, $\varepsilon_{m\eta}$, $\varepsilon_{N\lambda}$, $\varepsilon_{N\eta}$, and $\varepsilon_{\lambda\eta}$, for different nuclei and for some specific $\alpha$ values, in Tables \ref{ratios-table1}-\ref{ratios-table3}. Note that the interference coefficients $\varepsilon_{m\lambda}$ and $\varepsilon_{N\lambda}$ are equal. Using Eqs. (\ref{C_m}), (\ref{C_lambda}) and (\ref{C_m-lambda}) we see that, 
\begin{align}
\varepsilon_{m\lambda}=\varepsilon_{N\lambda}=\frac{\sqrt{\alpha}}{1+\alpha}\frac{\abs{G_{03}\mathcal{M}_{2-}-G_{04}\mathcal{M}_{1+}}}{\sqrt{G_{01}\abs{C_{\lambda}}}}.
\end{align}
Similarly, using \eq{C_N}, (\ref{C_lambda}) and (\ref{C_N-lambda}) we get,
\begin{align}
&\varepsilon_{m\eta}=\varepsilon_{N\eta}\non &=\frac{\sqrt{\alpha}}{1+\alpha}\frac{\abs{G_{03}\mathcal{M}_{2+}-G_{04}\mathcal{M}_{1-}-G_{05}M_P+G_{06}M_R}}{\sqrt{G_{01}\abs{C_{\eta}}}}.
\end{align}
Using \eq{C_m}, (\ref{C_N}) and (\ref{C_m-N}), one sees from \eq{epsilon} that the interference coefficient between the mass-mechanism and heavy neutrino exchange mechanism ($\varepsilon_{mN}$) is $\propto 2G^{\prime}_{01}/G_{01}$, which was considered in \rf{Ahmed:2017pqa}. We observe that maximum interference occurs for $\alpha=1$, i.e., when the pairs of individual mechanisms are equal to each other. Moreover, $\varepsilon_{ij}(\alpha)$ and $\varepsilon_{ij}(1/\alpha)$ are the same, as one can verify from \eq{epsilon}.

%%%%%%%%%%%%%%%%%%%%%%%%%%%%%%%%%%%%%%%%%%%%%%%%%%%%%%%%%%%%%%%%%%%%%%%%%%%%%%%%%%%%%%%%%%%%%%%%%%%%
%\vspace{-5mm}
\section{Results and Discussion}\label{Result}
From Tables \ref{ratios-table1}-\ref{ratios-table3} we observe an interference coefficient no larger than $\sim 18\%$ for the interference between the mass-mechanism and the $\lambda$ process ($\varepsilon_{m\lambda}$). The same conclusions can be drawn for the case for interference between RH-heavy neutrino exchange and the $\lambda$ mechanism ($\varepsilon_{N\lambda}=\varepsilon_{m\lambda}$). The interference coefficient for $\lambda$ and $\eta$ mechanisms is negligible with a maximum of $0.59\%$ for $^{82}$Se. The interference between the mass-mechanism and heavy neutrino exchange mechanism, $\varepsilon_{mN}$, was considered in \rf{Ahmed:2017pqa} for $\alpha=1$, see Eq. (25) and Table 1 of \cite{Ahmed:2017pqa}. For the interference between the mass mechanism and the $\eta$ mechanism, the maximum interference coefficient  ($\varepsilon_{m\eta}(\alpha=1)$) ranges between $30\%$ to $50\%$ with a maximum of about $49\%$ for $^{76}$Ge. The interference coefficient for RH-heavy neutrino exchange and the $\eta$ mechanism, $\varepsilon_{N\eta}$, has the same values.

We plot the coefficients for maximum interference, $\varepsilon_{m\lambda}(1)$, $\varepsilon_{m\eta}(1)$, and $\varepsilon_{\lambda\eta}(1)$ as functions of $Q$ value ($Q_{\beta\beta}$) of various nuclei in \fg{m-lambda}-\ref{lambda-eta}, respectively. We observe that $\varepsilon_{m\lambda}(1)$ and $\varepsilon_{m\eta}(1)$ decrease with $Q_{\beta\beta}$. In our study of the interference between the mass mechanism (\fg{Left-Light}) and the heavy-neutrino exchange for purely RH currents (\fg{Right-Heavy}) in \rf{Ahmed:2017pqa} we found a similar dependence of $\varepsilon_{mN}$ on $Q_{\beta\beta}$ (see Fig. 2 of \cite{Ahmed:2017pqa}). For the $\varepsilon_{\lambda\eta}(1)$ in \fg{lambda-eta} we do not observe any particular dependence on $Q_{\beta\beta}$.

In summary, we studied the contributions of the interference effects to the $\NDBD$ decay rate for four competing mechanisms arising from LRSM: (i) the regular mass mechanism for light-neutrino exchange of purely LH currents ($\eta_m$), (ii) the heavy-neutrino exchange mechanism for purely RH currents ($\eta_{N}$), (iii) the $\lambda$ mechanism ($\eta_{\lambda}$), and (iv) the $\eta$ mechanism ($\eta_{\eta}$). We extended our analysis of \rf{Ahmed:2017pqa} to interference effects between the `mass-mechanism' ($\eta_m$) and heavy-neutrino exchange mechanism ($\eta_N$) to the other five contributions. Besides several BSM scenarios, the LRSM is being actively investigated at the LHC \cite{Khachatryan:2014dka}. Several competing mechanisms have been proposed to contribute to $\NDBD$. It is important to know if different mechanisms can be disentangled. To that goal, analyzing the contribution of interference terms to the decay rate is essential. By comparing the decay rate of several nuclei of experimental interest one may be able to differentiate between two competing mechanisms, provided that the contribution of interference term is negligible \cite{PhysRevC.98.035502, PhysRevD.93.113014}. In the present study we have observed that most of the two-mechanisms interference terms introduce a relatively minor modification to the half-life, less than 20$\%$. However, the interference between the neutrino exchange mechanisms (light and heavy) and the $\eta$ mechanism are not small enough for the nuclei considered. In that case, the angular distribution of the emitted electrons can be used to distinguish between these two mechanisms, as has been discussed in \rf{PhysRevD.93.113014}. One should emphasize that the interference coefficients we found are not large enough to lead to a full cancellation of the decay rate (see \eq{gen-half-life}). Our conclusions are based on shell-model NME calculated with different sets of effective Hamiltonians and short-range correlation parametrizations, thus giving us confidence in their reliability.   

Our analysis of the interference terms in $\NDBD$ decay rate in the context of LRSM can also be extended for the EFT approach to $\NDBD$. Specifically, as discussed in Sec. \ref{EFT}, the interference of the amplitudes arising from scalar-pseudoscalar ($S\pm P$) and tensor ($T_R$) terms in $\mathcal{L}^{\text{\tiny{EFT}}}_{6}$ with the four amplitudes studied here. This analysis will be reported separately.                     

\vspace{-2mm}

\begin{acknowledgments}
Support from the U.S. Department of Energy Grant No. DE-SC0015376 is acknowledged. F.A. acknowledges the Science of Advanced Materials Program of Central Michigan University for their support through a research assistantship. F.A. would like to thank Andrei Neacsu for helpful discussions and for providing the source files for the NME and PSF values.
\end{acknowledgments}

\vspace{0mm}
\appendix*
\section{} \label{section-appendix}
\vspace{0mm}
In this appendix we tabulate the values of the ten PSF, $\{G_{01}-G_{09}$, $G^{\prime}_{01}\}$ and the thirteen NME, $\{M_F$, $M_{GT}$, $M_T$, $M_{F\omega}$, $M_{Fq}$, $M_{GT\omega}$, $M_{GTq}$, $M_{Tq}$, $M_P$, $M_R$, $M_{FN}$, $M_{GTN}\}$, $M_{TN}\}$ taken from the literature.
\vspace{0mm}
\begin{widetext}
\vspace{0mm}
\onecolumngrid
\begin{table}[b]
\vspace{0mm}
\begin{minipage}{\textwidth}
\centering
\caption{PSF in y$^{-1}$ for $0^+\to 0^+$ transition. Values of ($G_{01}-G_{09}$) are taken from \rf{PhysRevC.98.035502} for all the isotopes except for $^{124}$Sn. Values of $G^{\prime}_{01}$ are taken from \rf{Ahmed:2017pqa}.}
\begin{ruledtabular}
\begin{tabular}{lcccccc} \label{PSF-values}
			   			   			&$^{48}$Ca		&$^{76}$Ge		&$^{82}$Se		&$^{124}$Sn \cite{Neacsu:2016njp}		&$^{130}$Te		&$^{136}$Xe \\ 
\hline
$G_{01}\cdot 10^{14}$	   			&$2.45$ 			&$0.23$ 			&$0.10$			&$0.89$			&$1.41$			&$1.45$ \\

$G^{\prime}_{01}\cdot 10^{15}$ \cite{Ahmed:2017pqa}	   	&$1.09$ 			&$0.29$ 			&$0.76$			&$0.98$			&$1.37$			&$1.46$ \\

$G_{02}\cdot 10^{14}$	   			&$15.46$			&$0.35$ 			&$3.21$			&$1.68$			&$3.25$			&$3.15$ \\

$G_{03}\cdot 10^{14}$	   			&$1.82$			&$0.12$			&$0.65$			&$0.50$			&$0.85$			&$0.85$ \\

$G_{04}\cdot 10^{15}$	   			&$5.04$			&$0.42$			&$1.92$			&$1.56$			&$2.53$			&$2.58$ \\

$G_{05}\cdot 10^{13}$	   			&$3.28$ 			&$0.60$ 			&$2.16$			&$2.70$			&$4.12$			&$4.36$ \\

$G_{06}\cdot 10^{12}$	   			&$3.87$			&$0.50$			&$1.65$			&$1.47$			&$2.16$			&$2.21$ \\

$G_{07}\cdot 10^{10}$	   			&$2.85$			&$0.28$ 			&$1.20$			&$1.11$			&$1.75$			&$1.80$ \\

$G_{08}\cdot 10^{11}$	   			&$1.32$ 			&$0.17$ 			&$0.82$			&$1.04$			&$1.72$			&$1.83$ \\

$G_{09}\cdot 10^{10}$	   			&$15.55$			&$1.12$			&$4.42$			&$2.95$			&$4.47$			&$4.44$ \\
\end{tabular}
\end{ruledtabular}
\end{minipage}
\end{table}
\vspace{50mm}
\begin{table}[!htbp]
\begin{minipage}{\textwidth}
\caption{Dimensionless NME for $0^+\to 0^+$ transition. Values taken from \rf{PhysRevC.98.035502, PhysRevC.93.024308, Neacsu:2016njp}. From \rf{PhysRevC.98.035502}, we have considered the NME calculated with the CMU effective Hamiltonians and CD-Bonn SRC-parametrization.}
\begin{ruledtabular}
\begin{tabular}{lccccccc}\label{NME-values}

&               &$^{48}$Ca	&$^{76}$Ge 	&$^{82}$Se 	&$^{124}$Sn \cite{PhysRevC.93.024308, Neacsu:2016njp} 	&$^{130}$Te 	&$^{136}$Xe \\ 
\hline 

&$M_{GT}$ 	    &$-0.805$   		&$-3.200$  		&$-3.000$		&$-1.853$		&$-1.658$  		&$1.501$  \\ 

&$M_{F}$        &$0.233$    		&$0.674$   		&$0.632$   		&$0.467$			&$0.438$   		&$-0.400$  \\

&$M_{T}$        &$-0.073$   		&$-0.011$  		&$-0.012$  		&$-0.019$		&$0.006$   		&$-0.007$  \\

&$M_{GTN}$ 	    &$-55.890$   	&$-156.493$  	&$-144.907$  	&$-113.364$ 		&$-103.025$		&$92.565$  \\ 

&$M_{FN}$       &$22.893$    	&$62.649$   		&$58.091$   		&$43.295$   		&$40.984$	  	&$-36.942$ \\

&$M_{TN}$       &$-11.308$   	&$-0.205$  		&$-0.513$  		&$-3.827$   		&$2.022$  		&$-2.178$ \\

&$M_{GTq}$      &$-0.709$   		&$-3.228$  		&$-3.034$  		&$1.793$			&$-1.587$  		&$1.440$   \\

&$M_{Fq}$       &$0.121$    		&$0.383$   		&$0.362$   		&$-0.267$		&$0.249$   		&$-0.230$  \\

&$M_{Tq}$       &$0.173$    		&$0.059$   		&$0.058$   		&$0.011$ 		&$0.013$   		&$-0.012$  \\

&$M_{GT\omega}$ &$-0.930$   		&$-3.501$  		&$-3.287$  		&$2.053$			&$-1.855$  		&$1.682$   \\

&$M_{F\omega}$  &$0.232$   	 	&$0.659$   		&$0.618$   		&$-0.456$		&$0.427$   		&$-0.391$  \\

&$M_{R}$        &$-1.001$   		&$-3.243$  		&$-3.088$  		&$2.663$			&$-2.530$  		&$2.312$   \\

&$M_{P}$        &$-0.390$   		&$2.435$   		&$2.303$   		&$-2.060$		&$1.707$   		&$-1.600$  \\
\end{tabular}
\end{ruledtabular}
\end{minipage}
\end{table}
\end{widetext}

\twocolumngrid
\vspace{70mm}

\bibliographystyle{apsrev}
\bibliography{Ref-LRSM_Intf}

\begin{thebibliography}{36}
\expandafter\ifx\csname natexlab\endcsname\relax\def\natexlab#1{#1}\fi
\expandafter\ifx\csname bibnamefont\endcsname\relax
  \def\bibnamefont#1{#1}\fi
\expandafter\ifx\csname bibfnamefont\endcsname\relax
  \def\bibfnamefont#1{#1}\fi
\expandafter\ifx\csname citenamefont\endcsname\relax
  \def\citenamefont#1{#1}\fi
\expandafter\ifx\csname url\endcsname\relax
  \def\url#1{\texttt{#1}}\fi
\expandafter\ifx\csname urlprefix\endcsname\relax\def\urlprefix{URL }\fi
\providecommand{\bibinfo}[2]{#2}
\providecommand{\eprint}[2][]{\url{#2}}

\bibitem[{\citenamefont{Schechter and Valle}(1982)}]{PhysRevD.25.2951}
\bibinfo{author}{\bibfnamefont{J.}~\bibnamefont{Schechter}} \bibnamefont{and}
  \bibinfo{author}{\bibfnamefont{J.~W.~F.} \bibnamefont{Valle}},
  \bibinfo{journal}{Phys. Rev. D} \textbf{\bibinfo{volume}{25}},
  \bibinfo{pages}{2951} (\bibinfo{year}{1982}),
  \urlprefix\url{http://link.aps.org/doi/10.1103/PhysRevD.25.2951}.

\bibitem[{\citenamefont{Horoi}(2013)}]{PhysRevC.87.014320}
\bibinfo{author}{\bibfnamefont{M.}~\bibnamefont{Horoi}},
  \bibinfo{journal}{Phys. Rev. C} \textbf{\bibinfo{volume}{87}},
  \bibinfo{pages}{014320} (\bibinfo{year}{2013}),
  \urlprefix\url{http://link.aps.org/doi/10.1103/PhysRevC.87.014320}.

\bibitem[{\citenamefont{Vergados et~al.}(2012)\citenamefont{Vergados, Ejiri,
  and Simkovic}}]{0034-4885-75-10-106301}
\bibinfo{author}{\bibfnamefont{J.~D.} \bibnamefont{Vergados}},
  \bibinfo{author}{\bibfnamefont{H.}~\bibnamefont{Ejiri}}, \bibnamefont{and}
  \bibinfo{author}{\bibfnamefont{F.}~\bibnamefont{Simkovic}},
  \bibinfo{journal}{Rept. Prog. Phys.} \textbf{\bibinfo{volume}{75}},
  \bibinfo{pages}{106301} (\bibinfo{year}{2012}), \eprint{1205.0649}.

\bibitem[{\citenamefont{Khachatryan et~al.}(2014)}]{Khachatryan:2014dka}
\bibinfo{author}{\bibfnamefont{V.}~\bibnamefont{Khachatryan}}
  \bibnamefont{et~al.} (\bibinfo{collaboration}{CMS}), \bibinfo{journal}{Eur.
  Phys. J. C} \textbf{\bibinfo{volume}{74}}, \bibinfo{pages}{3149}
  (\bibinfo{year}{2014}), \eprint{1407.3683}.

\bibitem[{\citenamefont{Barry and Rodejohann}(2013)}]{Barry:2013xxa}
\bibinfo{author}{\bibfnamefont{J.}~\bibnamefont{Barry}} \bibnamefont{and}
  \bibinfo{author}{\bibfnamefont{W.}~\bibnamefont{Rodejohann}},
  \bibinfo{journal}{JHEP} \textbf{\bibinfo{volume}{09}}, \bibinfo{pages}{153}
  (\bibinfo{year}{2013}), \eprint{1303.6324}.

\bibitem[{\citenamefont{Mohapatra and Senjanovic}(1980)}]{PhysRevLett.44.912}
\bibinfo{author}{\bibfnamefont{R.~N.} \bibnamefont{Mohapatra}}
  \bibnamefont{and}
  \bibinfo{author}{\bibfnamefont{G.}~\bibnamefont{Senjanovic}},
  \bibinfo{journal}{Phys. Rev. Lett.} \textbf{\bibinfo{volume}{44}},
  \bibinfo{pages}{912} (\bibinfo{year}{1980}),
  \urlprefix\url{http://link.aps.org/doi/10.1103/PhysRevLett.44.912}.

\bibitem[{\citenamefont{Mohapatra and
  Senjanovic}(1981)}]{Mohapatra+Senjanovic1981}
\bibinfo{author}{\bibfnamefont{R.~N.} \bibnamefont{Mohapatra}}
  \bibnamefont{and}
  \bibinfo{author}{\bibfnamefont{G.}~\bibnamefont{Senjanovic}},
  \bibinfo{journal}{Phys. Rev. D} \textbf{\bibinfo{volume}{23}},
  \bibinfo{pages}{165} (\bibinfo{year}{1981}),
  \urlprefix\url{http://link.aps.org/doi/10.1103/PhysRevD.23.165}.

\bibitem[{\citenamefont{Faessler et~al.}(2011)\citenamefont{Faessler, Meroni,
  Petcov, Simkovic, and Vergados}}]{PhysRevD.83.113003}
\bibinfo{author}{\bibfnamefont{A.}~\bibnamefont{Faessler}},
  \bibinfo{author}{\bibfnamefont{A.}~\bibnamefont{Meroni}},
  \bibinfo{author}{\bibfnamefont{S.~T.} \bibnamefont{Petcov}},
  \bibinfo{author}{\bibfnamefont{F.}~\bibnamefont{Simkovic}}, \bibnamefont{and}
  \bibinfo{author}{\bibfnamefont{J.}~\bibnamefont{Vergados}},
  \bibinfo{journal}{Phys. Rev. D} \textbf{\bibinfo{volume}{83}},
  \bibinfo{pages}{113003} (\bibinfo{year}{2011}),
  \urlprefix\url{http://link.aps.org/doi/10.1103/PhysRevD.83.113003}.

\bibitem[{\citenamefont{Ahmed et~al.}(2017)\citenamefont{Ahmed, Neacsu, and
  Horoi}}]{Ahmed:2017pqa}
\bibinfo{author}{\bibfnamefont{F.}~\bibnamefont{Ahmed}},
  \bibinfo{author}{\bibfnamefont{A.}~\bibnamefont{Neacsu}}, \bibnamefont{and}
  \bibinfo{author}{\bibfnamefont{M.}~\bibnamefont{Horoi}},
  \bibinfo{journal}{Phys. Lett. B} \textbf{\bibinfo{volume}{769}},
  \bibinfo{pages}{299} (\bibinfo{year}{2017}), \eprint{1701.03177}.

\bibitem[{\citenamefont{Pati and Salam}(1974)}]{Pati:1974yy}
\bibinfo{author}{\bibfnamefont{J.~C.} \bibnamefont{Pati}} \bibnamefont{and}
  \bibinfo{author}{\bibfnamefont{A.}~\bibnamefont{Salam}},
  \bibinfo{journal}{Phys. Rev. D} \textbf{\bibinfo{volume}{10}},
  \bibinfo{pages}{275} (\bibinfo{year}{1974}), \bibinfo{note}{[Erratum: Phys.
  Rev.D11,703(1975)]}.

\bibitem[{\citenamefont{Mohapatra and Pati}(1975)}]{Mohapatra:1974gc}
\bibinfo{author}{\bibfnamefont{R.~N.} \bibnamefont{Mohapatra}}
  \bibnamefont{and} \bibinfo{author}{\bibfnamefont{J.~C.} \bibnamefont{Pati}},
  \bibinfo{journal}{Phys. Rev. D} \textbf{\bibinfo{volume}{11}},
  \bibinfo{pages}{2558} (\bibinfo{year}{1975}).

\bibitem[{\citenamefont{Bhupal~Dev et~al.}(2015)\citenamefont{Bhupal~Dev,
  Goswami, and Mitra}}]{Dev:2014xea}
\bibinfo{author}{\bibfnamefont{P.~S.} \bibnamefont{Bhupal~Dev}},
  \bibinfo{author}{\bibfnamefont{S.}~\bibnamefont{Goswami}}, \bibnamefont{and}
  \bibinfo{author}{\bibfnamefont{M.}~\bibnamefont{Mitra}},
  \bibinfo{journal}{Phys. Rev. D} \textbf{\bibinfo{volume}{91}},
  \bibinfo{pages}{113004} (\bibinfo{year}{2015}), \eprint{1405.1399}.

\bibitem[{\citenamefont{Borah et~al.}(2018)\citenamefont{Borah, Dasgupta, and
  Patra}}]{Borah:2017ldt}
\bibinfo{author}{\bibfnamefont{D.}~\bibnamefont{Borah}},
  \bibinfo{author}{\bibfnamefont{A.}~\bibnamefont{Dasgupta}}, \bibnamefont{and}
  \bibinfo{author}{\bibfnamefont{S.}~\bibnamefont{Patra}},
  \bibinfo{journal}{Int. J. Mod. Phys. A} \textbf{\bibinfo{volume}{33}},
  \bibinfo{pages}{1850198} (\bibinfo{year}{2018}), \eprint{1706.02456}.

\bibitem[{\citenamefont{Burgess and Moore}(2006)}]{Burgess:2007zi}
\bibinfo{author}{\bibfnamefont{C.~P.} \bibnamefont{Burgess}} \bibnamefont{and}
  \bibinfo{author}{\bibfnamefont{G.~D.} \bibnamefont{Moore}},
  \emph{\bibinfo{title}{{The Standard Model: A Primer}}}
  (\bibinfo{publisher}{Cambridge University Press}, \bibinfo{year}{2006}), ISBN
  \bibinfo{isbn}{9780511254857}.

\bibitem[{\citenamefont{Grimus}(1993)}]{Grimus:1993fx}
\bibinfo{author}{\bibfnamefont{W.}~\bibnamefont{Grimus}}, in
  \emph{\bibinfo{booktitle}{{Elementary particle physics. Proceedings, 4th
  Hellenic School, Corfu, Greece, September 2-20, 1992. 1\&2}}}
  (\bibinfo{year}{1993}), pp. \bibinfo{pages}{619--632},
  \urlprefix\url{https://inis.iaea.org/collection/NCLCollectionStore/_Public/25/012/25012246.pdf}.

\bibitem[{\citenamefont{Senjanovic and Tello}(2016)}]{Senjanovic:2015yea}
\bibinfo{author}{\bibfnamefont{G.}~\bibnamefont{Senjanovic}} \bibnamefont{and}
  \bibinfo{author}{\bibfnamefont{V.}~\bibnamefont{Tello}},
  \bibinfo{journal}{Phys. Rev. D} \textbf{\bibinfo{volume}{94}},
  \bibinfo{pages}{095023} (\bibinfo{year}{2016}), \eprint{1502.05704}.

\bibitem[{\citenamefont{Doi et~al.}(1985)\citenamefont{Doi, Kotani, and
  Takasugi}}]{Doi+Kotani1985}
\bibinfo{author}{\bibfnamefont{M.}~\bibnamefont{Doi}},
  \bibinfo{author}{\bibfnamefont{T.}~\bibnamefont{Kotani}}, \bibnamefont{and}
  \bibinfo{author}{\bibfnamefont{E.}~\bibnamefont{Takasugi}},
  \bibinfo{journal}{Prog. Theor. Phys. Suppl.} \textbf{\bibinfo{volume}{83}},
  \bibinfo{pages}{1} (\bibinfo{year}{1985}),
  \urlprefix\url{http://ptps.oxfordjournals.org/content/83/1.abstract}.

\bibitem[{\citenamefont{Hirsch and
  Klapdor-Kleingrothaus}(1996)}]{Hirsch:1995rf}
\bibinfo{author}{\bibfnamefont{M.}~\bibnamefont{Hirsch}} \bibnamefont{and}
  \bibinfo{author}{\bibfnamefont{H.~V.} \bibnamefont{Klapdor-Kleingrothaus}},
  in \emph{\bibinfo{booktitle}{{Double beta decay and related topics.
  Proceedings of the International Workshop held at European Center for
  Theoretical Studies (ECT*)}}} (\bibinfo{publisher}{World Scientific},
  \bibinfo{year}{1996}), pp. \bibinfo{pages}{175--191}.

\bibitem[{\citenamefont{Srednicki}(2007)}]{Srednicki2007}
\bibinfo{author}{\bibfnamefont{M.}~\bibnamefont{Srednicki}},
  \emph{\bibinfo{title}{{Quantum field theory}}} (\bibinfo{publisher}{Cambridge
  University Press}, \bibinfo{year}{2007}), ISBN \bibinfo{isbn}{9780521864497}.

\bibitem[{\citenamefont{Pas et~al.}(1999)\citenamefont{Pas, Hirsch,
  Klapdor-Kleingrothaus, and Kovalenko}}]{PAS1999194}
\bibinfo{author}{\bibfnamefont{H.}~\bibnamefont{Pas}},
  \bibinfo{author}{\bibfnamefont{H.}~\bibnamefont{Hirsch}},
  \bibinfo{author}{\bibfnamefont{H.~V.} \bibnamefont{Klapdor-Kleingrothaus}},
  \bibnamefont{and} \bibinfo{author}{\bibfnamefont{S.~G.}
  \bibnamefont{Kovalenko}}, \bibinfo{journal}{Phys. Lett. B}
  \textbf{\bibinfo{volume}{453}}, \bibinfo{pages}{194 } (\bibinfo{year}{1999}),
  ISSN \bibinfo{issn}{0370-2693},
  \urlprefix\url{http://www.sciencedirect.com/science/article/pii/S0370269399003305}.

\bibitem[{\citenamefont{Barry et~al.}(2014)\citenamefont{Barry, Heeck, and
  Rodejohann}}]{Barry:2014ika}
\bibinfo{author}{\bibfnamefont{J.}~\bibnamefont{Barry}},
  \bibinfo{author}{\bibfnamefont{J.}~\bibnamefont{Heeck}}, \bibnamefont{and}
  \bibinfo{author}{\bibfnamefont{W.}~\bibnamefont{Rodejohann}},
  \bibinfo{journal}{JHEP} \textbf{\bibinfo{volume}{07}}, \bibinfo{pages}{081}
  (\bibinfo{year}{2014}), \eprint{1404.5955}.

\bibitem[{\citenamefont{Tello et~al.}(2011)\citenamefont{Tello, Nemevsek,
  Nesti, Senjanovic, and Vissani}}]{Tello:2010am}
\bibinfo{author}{\bibfnamefont{V.}~\bibnamefont{Tello}},
  \bibinfo{author}{\bibfnamefont{M.}~\bibnamefont{Nemevsek}},
  \bibinfo{author}{\bibfnamefont{F.}~\bibnamefont{Nesti}},
  \bibinfo{author}{\bibfnamefont{G.}~\bibnamefont{Senjanovic}},
  \bibnamefont{and} \bibinfo{author}{\bibfnamefont{F.}~\bibnamefont{Vissani}},
  \bibinfo{journal}{Phys. Rev. Lett.} \textbf{\bibinfo{volume}{106}},
  \bibinfo{pages}{151801} (\bibinfo{year}{2011}), \eprint{1011.3522}.

\bibitem[{\citenamefont{Stoica and Mirea}(2013)}]{PhysRevC.88.037303}
\bibinfo{author}{\bibfnamefont{S.}~\bibnamefont{Stoica}} \bibnamefont{and}
  \bibinfo{author}{\bibfnamefont{M.}~\bibnamefont{Mirea}},
  \bibinfo{journal}{Phys. Rev. C} \textbf{\bibinfo{volume}{88}},
  \bibinfo{pages}{037303} (\bibinfo{year}{2013}),
  \urlprefix\url{https://link.aps.org/doi/10.1103/PhysRevC.88.037303}.

\bibitem[{\citenamefont{Neacsu and Horoi}(2016{\natexlab{a}})}]{Neacsu:2015uja}
\bibinfo{author}{\bibfnamefont{A.}~\bibnamefont{Neacsu}} \bibnamefont{and}
  \bibinfo{author}{\bibfnamefont{M.}~\bibnamefont{Horoi}},
  \bibinfo{journal}{Adv. High Energy Phys.} \textbf{\bibinfo{volume}{2016}},
  \bibinfo{pages}{7486712} (\bibinfo{year}{2016}{\natexlab{a}}),
  \eprint{1510.00882}.

\bibitem[{\citenamefont{Suhonen and Civitarese}(1998)}]{SuhonenCivitarese1998}
\bibinfo{author}{\bibfnamefont{J.}~\bibnamefont{Suhonen}} \bibnamefont{and}
  \bibinfo{author}{\bibfnamefont{O.}~\bibnamefont{Civitarese}},
  \bibinfo{journal}{Phys. Rep.} \textbf{\bibinfo{volume}{300}},
  \bibinfo{pages}{123} (\bibinfo{year}{1998}).

\bibitem[{\citenamefont{Doi and Kotani}(1993)}]{10.1143/ptp/89.1.139}
\bibinfo{author}{\bibfnamefont{M.}~\bibnamefont{Doi}} \bibnamefont{and}
  \bibinfo{author}{\bibfnamefont{T.}~\bibnamefont{Kotani}},
  \bibinfo{journal}{"Prog. Theor. Phys."} \textbf{\bibinfo{volume}{89}},
  \bibinfo{pages}{139} (\bibinfo{year}{1993}),
  \eprint{https://academic.oup.com/ptp/article-pdf/89/1/139/5207768/89-1-139.pdf},
  \urlprefix\url{https://doi.org/10.1143/ptp/89.1.139}.

\bibitem[{\citenamefont{Horoi and Neacsu}(2018)}]{PhysRevC.98.035502}
\bibinfo{author}{\bibfnamefont{M.}~\bibnamefont{Horoi}} \bibnamefont{and}
  \bibinfo{author}{\bibfnamefont{A.}~\bibnamefont{Neacsu}},
  \bibinfo{journal}{Phys. Rev. C} \textbf{\bibinfo{volume}{98}},
  \bibinfo{pages}{035502} (\bibinfo{year}{2018}).

\bibitem[{\citenamefont{Deppisch et~al.}(2012)\citenamefont{Deppisch, Hirsch,
  and Pas}}]{Deppisch2012}
\bibinfo{author}{\bibfnamefont{F.~F.} \bibnamefont{Deppisch}},
  \bibinfo{author}{\bibfnamefont{M.}~\bibnamefont{Hirsch}}, \bibnamefont{and}
  \bibinfo{author}{\bibfnamefont{H.}~\bibnamefont{Pas}}, \bibinfo{journal}{J.
  Phys. G} \textbf{\bibinfo{volume}{39}}, \bibinfo{pages}{124007}
  (\bibinfo{year}{2012}).

\bibitem[{\citenamefont{Cirigliano et~al.}(2018)\citenamefont{Cirigliano,
  Dekens, de~Vries, Graesser, and Mereghetti}}]{Cirigliano:2018yza}
\bibinfo{author}{\bibfnamefont{V.}~\bibnamefont{Cirigliano}},
  \bibinfo{author}{\bibfnamefont{W.}~\bibnamefont{Dekens}},
  \bibinfo{author}{\bibfnamefont{J.}~\bibnamefont{de~Vries}},
  \bibinfo{author}{\bibfnamefont{M.~L.} \bibnamefont{Graesser}},
  \bibnamefont{and}
  \bibinfo{author}{\bibfnamefont{E.}~\bibnamefont{Mereghetti}},
  \bibinfo{journal}{JHEP} \textbf{\bibinfo{volume}{12}}, \bibinfo{pages}{097}
  (\bibinfo{year}{2018}), \eprint{1806.02780}.

\bibitem[{\citenamefont{del Aguila et~al.}(2012)\citenamefont{del Aguila,
  Aparici, Bhattacharya, Santamaria, and Wudka}}]{delAguila:2012nu}
\bibinfo{author}{\bibfnamefont{F.}~\bibnamefont{del Aguila}},
  \bibinfo{author}{\bibfnamefont{A.}~\bibnamefont{Aparici}},
  \bibinfo{author}{\bibfnamefont{S.}~\bibnamefont{Bhattacharya}},
  \bibinfo{author}{\bibfnamefont{A.}~\bibnamefont{Santamaria}},
  \bibnamefont{and} \bibinfo{author}{\bibfnamefont{J.}~\bibnamefont{Wudka}},
  \bibinfo{journal}{JHEP} \textbf{\bibinfo{volume}{06}}, \bibinfo{pages}{146}
  (\bibinfo{year}{2012}), \eprint{1204.5986}.

\bibitem[{\citenamefont{Horoi and Stoica}(2010)}]{HoroiStoica2010}
\bibinfo{author}{\bibfnamefont{M.}~\bibnamefont{Horoi}} \bibnamefont{and}
  \bibinfo{author}{\bibfnamefont{S.}~\bibnamefont{Stoica}},
  \bibinfo{journal}{Phys. Rev. C} \textbf{\bibinfo{volume}{81}},
  \bibinfo{pages}{024321} (\bibinfo{year}{2010}).

\bibitem[{\citenamefont{Neacsu and Horoi}(2015)}]{NeacsuHoroi2015}
\bibinfo{author}{\bibfnamefont{A.}~\bibnamefont{Neacsu}} \bibnamefont{and}
  \bibinfo{author}{\bibfnamefont{M.}~\bibnamefont{Horoi}},
  \bibinfo{journal}{Phys. Rev. C} \textbf{\bibinfo{volume}{91}},
  \bibinfo{pages}{024309} (\bibinfo{year}{2015}).

\bibitem[{\citenamefont{Horoi and
  Neacsu}(2016{\natexlab{a}})}]{PhysRevC.93.024308}
\bibinfo{author}{\bibfnamefont{M.}~\bibnamefont{Horoi}} \bibnamefont{and}
  \bibinfo{author}{\bibfnamefont{A.}~\bibnamefont{Neacsu}},
  \bibinfo{journal}{Phys. Rev. C} \textbf{\bibinfo{volume}{93}},
  \bibinfo{pages}{024308} (\bibinfo{year}{2016}{\natexlab{a}}),
  \urlprefix\url{https://link.aps.org/doi/10.1103/PhysRevC.93.024308}.

\bibitem[{\citenamefont{Stefanik et~al.}(2015)\citenamefont{Stefanik,
  Dvornicky, Simkovic, and Vogel}}]{Stefanik:2015twa}
\bibinfo{author}{\bibfnamefont{D.}~\bibnamefont{Stefanik}},
  \bibinfo{author}{\bibfnamefont{R.}~\bibnamefont{Dvornicky}},
  \bibinfo{author}{\bibfnamefont{F.}~\bibnamefont{Simkovic}}, \bibnamefont{and}
  \bibinfo{author}{\bibfnamefont{P.}~\bibnamefont{Vogel}},
  \bibinfo{journal}{Phys. Rev. C} \textbf{\bibinfo{volume}{92}},
  \bibinfo{pages}{055502} (\bibinfo{year}{2015}), \eprint{1506.07145}.

\bibitem[{\citenamefont{Horoi and
  Neacsu}(2016{\natexlab{b}})}]{PhysRevD.93.113014}
\bibinfo{author}{\bibfnamefont{M.}~\bibnamefont{Horoi}} \bibnamefont{and}
  \bibinfo{author}{\bibfnamefont{A.}~\bibnamefont{Neacsu}},
  \bibinfo{journal}{Phys. Rev. D} \textbf{\bibinfo{volume}{93}},
  \bibinfo{pages}{113014} (\bibinfo{year}{2016}{\natexlab{b}}),
  \urlprefix\url{https://link.aps.org/doi/10.1103/PhysRevD.93.113014}.

\bibitem[{\citenamefont{Neacsu and Horoi}(2016{\natexlab{b}})}]{Neacsu:2016njp}
\bibinfo{author}{\bibfnamefont{A.}~\bibnamefont{Neacsu}} \bibnamefont{and}
  \bibinfo{author}{\bibfnamefont{M.}~\bibnamefont{Horoi}},
  \bibinfo{journal}{Adv. High Energy Phys.} \textbf{\bibinfo{volume}{2016}},
  \bibinfo{pages}{1903767} (\bibinfo{year}{2016}{\natexlab{b}}),
  \eprint{1607.01295}.

\end{thebibliography}

\end{document}